\documentclass[twocolumn,showpacs,aps,pra,eqsecnum]{revtex4-1}

\usepackage{amsmath}
\usepackage{amssymb}
\usepackage{epsfig}

\usepackage{color}

\begin{document}

%\draft

\title{Quantum speed of evolution in a  Markovian bosonic environment}
\author{Paulina Marian }
\email{paulina.marian@g.unibuc.ro}
\author{Tudor A. Marian}
\email{tudor.marian@g.unibuc.ro}
\affiliation{ Centre for Advanced  Quantum Physics,
Department of Physics, University of Bucharest, 
R-077125 M\u{a}gurele, Romania}

\date{\today}

\begin{abstract}
We present explicit evaluations of quantum speed limit times pertinent to the Markovian 
dynamics of an open continuous-variable system. Specifically, we consider the standard 
setting of a cavity mode of the quantum radiation field weakly coupled to a thermal bosonic 
reservoir. The evolution of the field state is ruled by the quantum optical master equation, 
which is known to have an exact analytic solution. Starting from a pure input state, 
we employ two indicators of how different the initial and evolved states are, namely, 
the fidelity of evolution and the Hilbert-Schmidt distance of evolution. The former 
was introduced by del Campo {\em et al.} [Phys. Rev. Lett. {\bf 110}, 050403 (2013)], 
who derived a time-independent speed limit for the evolution of a Markovian open system.
We evaluate it for this field-reservoir setting, with an arbitrary input pure state of the field mode. 
The resultant formula is then specialized to the coherent and Fock states. On the other hand, 
we exploit an alternative approach that employs both indicators of evolution mentioned above. 
Their rates of change have the same upper bound, and consequently provide a unique 
time-dependent quantum speed limit. It turns out that the associate quantum speed 
limit time built with the Hilbert-Schmidt metric is tighter than the fidelity-based one. 
As apposite applications, we investigate the damping of the coherent and Fock states 
by using the characteristic functions
of the corresponding evolved states. General expressions of both the fidelity and the Hilbert-Schmidt 
distance of evolution are obtained and analyzed for these two classes of input states. In the case of a coherent state, we derive accurate 
formulas for their common speed limit and the pair of associate limit times. We also find 
exact expressions of the same quantities in the limiting case of thermalization of the vacuum 
state, as well as for dissipation of one- and two-photon states.  
\end{abstract}
%\pacs{42.50.Dv;03.65.Yz}
\maketitle

\section{Introduction}

The ingenious derivation of a time-energy uncertainty relation by Mandelstam 
and Tamm \cite{MT}  can be considered the starting point of an intense research regarding 
a fundamental limit imposed by quantum mechanics  on the minimal evolution time between 
two distinguishable states of an isolated quantum system. Such a system has 
a time-independent Hamiltonian $\hat H$. Mandelstam and Tamm found that its time 
of evolution between two orthogonal pure states, denoted $t_{\bot}$, has a lower bound 
which is inversely proportional to the standard deviation of the energy, 
$\Delta E= \sqrt{\langle{\hat H}^2\rangle- \langle{\hat H}\rangle^2}$\, : 
\begin{equation}
t_{\bot} \geqq {\tau}^{\rm (MT)}_{\bot}=\frac{\pi}{2} \frac{\hbar} {\Delta E}.
\label{MT}
\end{equation}
This inequality is related to that presented in many textbooks, for instance 
in Ref. \cite{Messiah1}, as the time-energy uncertainty relation: 
$T \Delta E \geqq \frac{\hbar}{2}$. Here $T$ denotes a conventional time of evolution 
of an isolated quantum system, which is equal to the minimal characteristic 
evolution time of any time-independent observable. However,  
${\tau}^{\rm (MT)}_{\bot}$ has a simpler meaning,  because it is defined 
in terms of the non-decay probability over a given time interval $[0, \, t]$. 
By scrutinizing the Mandelstam-Tamm proof,  Bhattacharyya \cite{Bhatta} recovered 
Eq. (\ref{MT}), as well as a lower bound for the half-life of a non-stationary state 
of an isolated quantum system:
$$t_{1/2}\geqq {\tau}^{\rm (MT)}_{1/2}=\frac{\pi}{4} \frac{\hbar} {\Delta E}.$$

More recently,  Margolus and Levitin \cite{ML} found an alternative lower bound 
to the orthogonalization time $t_{\bot}$ that is inversely proportional to the mean
excitation energy,  i. e., the difference between the average energy 
$E=\langle \hat H \rangle$ and the ground-state energy $E_g=0$:
\begin{equation}
t_{\bot} \geqq {\tau}^{\rm (ML)}_{\bot}=\frac{\pi}{2} \, \frac{\hbar} {E}.
\label{ML}
\end{equation}
The elegant Margolus-Levitin derivation of the bound (\ref{ML}) was later generalized 
in a versatile way that allows one to retrieve the Mandelstam-Tamm bound (\ref{MT}) 
to the orthogonalization time $t_{\bot}$ as another distinct particular case \cite{KZ}.  
Obviously, the actual limit time of orthogonalization ${\tau}_{\bot}$
is the largest one from the pair of bounds (\ref{MT}) and (\ref{ML}):
\begin{align}
& {\tau}_{\bot}={\rm max}\left\{ {\tau}^{\rm (MT)}_{\bot}, \, {\tau}^{\rm (ML)}_{\bot} \right\}   
\notag    \\
& =\frac{\pi \hbar} {E+\Delta{E}-|E-\Delta{E}| }.
\label{LT}
\end{align}
Although the bound (\ref{LT}) can be attained only if $\Delta{E}=E$, it remains tight 
for any values of $E$ and $\Delta{E}$ \cite{LT}.

The classic paper by Anandan and Aharonov \cite{AA} is devoted to the geometry
underlying the unitary evolution of a closed quantum system.. The authors proved 
that the Fubini-Study distance $s(t)$ between an initial pure state and the evolved one 
along a suitable path ${\cal C}$ in the projective Hilbert space has the velocity of change 
\begin{equation}
{\dot s}(t)=\frac{ {\Delta E}(t) }{\hbar}.
\label{dotFS}
\end{equation}
By using the time-averaged standard deviation
$$ \overline{\Delta E}(t):=\frac{1}{t} \int_{0 \, ( {\cal C} ) }^{t} dt^{\prime} {\Delta E}(t^{\prime}), $$
Eq. (\ref{dotFS}) provides the Fubini-Study evolution distance along the path ${\cal C}$:
\begin{equation}
s(t)=t \, \frac{\overline{\Delta E}(t) }{\hbar}.
\label{FS}
\end{equation}

An insightful application concerns the precession of the magnetic dipole moment $\mu$
of a spin-$\frac{1}{2}$ particle driven by a uniform, variable magnetic field, 
${\bf B}(t)=B(t){\bf e}_z$. This evolution is ruled by the Hamiltonian 
${\hat H}(t)=-\mu B(t) {\hat \sigma}_z$, acting on the Euclidean Hilbert space 
${\mathbb C}^2$. The associate projective Hilbert space ${\mathbb C}{\mathbb P}^1$
is diffeomorphic with the unit $2$-sphere ${\mathbb S}^2$ via the stereographic 
projection. Further, the Fubini-Study metric on ${\mathbb C}{\mathbb P}^1$ 
is proportional to the round metric on ${\mathbb S}^2$:
$$ {ds}^2=\frac{1}{4}\left[ {d \theta}^2 + {\sin}^2 (\theta) \, {d \phi}^2 \right], $$
where $\theta$ and $\phi$ are the  spherical polar angles. Note that two orthogonal
pure spin states are antipodal points on the Bloch sphere ${\mathbb S}^2$. 
By definition, any Fubini-Study distance  between them cannot be shorter 
than the geodesic one: $s( t_{\bot} ) \geqq s( {\tau}_{\bot} ):=\frac{\pi}{2}$. 
Taking into account Eq. (\ref{FS}), this inequality reads
\begin{equation}
t_{\bot} \geqq \frac{\pi}{2} \frac{\hbar} {\overline{\Delta E}(t) }
\label{AA}
\end{equation}
and becomes saturated when the spin precession about the magnetic field 
follows a geodesic. 

The formula (\ref{AA}) is twofold meaningful. First, this example affords a generalization 
of the Mandelstam-Tamm bound (\ref{MT}) to a closed quantum system whose evolution 
is driven by a time-dependent Hamiltonian. Accordingly, although the above-discussed 
bound becomes now time-dependent, there is no practical impediment to estimate
its tightness. Second, Eq. (\ref{AA}) reveals the deep connectionbetween the limit times 
for a unitary pure-state evolution and the Fubini-Study metric on the projective Hilbert 
space, when chosen as a measure of distinguishability between pure states.

Interesting analyses regarding  the unitary evolution of an isolated quantum system 
are made in Refs. \cite{V,U}. In these notes, two distinct straightforward methods 
are applied  to evaluate the maximal speed of the probability of transition from 
an initial pure state to an evolved one. The final result is a constant quantum velocity 
(\ref{dotFS}), which is equivalent to the Mandelstam-Tamm lower bound (\ref{MT}). 
In fact, Vaidman introduced the notion of quantum speed limit (QSL) for a pure-state 
unitary evolution \cite{V} as a necessary tool for obtaining what are currently called 
quantum speed limit times (QSLTs) \cite{GLM}. 

As shown by Poggi in a recent overview of quantum control times \cite{Poggi}, 
Eq. (\ref{FS}) encompasses the unitary evolution of a system driven towards 
a target state. The control field enters the Hamiltonian ${\hat H}(u(t))$ as a bounded 
vector function $u(t) \in {\mathbb R}^{n}, \, ||u(t)|| \leqq ||u||_{\rm max},$ 
whose time dependence is unknown. A fruitful application of Eq. (\ref{FS}) in order 
to obtain a lower bound on the minimum control time, which is independent 
of the actual form of the function $u(t)$, goes back to Pfeifer's letter \cite{Pfeifer}.

The innovative use in Ref. \cite{AA} of the Fubini-Study metric to characterize 
the pure-state evolution paved the way to an important extension to the case 
of mixed states treated in Ref. \cite{GLM}. Here, a general definition of the QSL 
to the evolution of a system towards a given quantum state (pure or mixed) 
relies on the generalization of the Fubini-Study metric for pure states 
to the Bures metric for mixed ones \cite{Bures}.  As such, the quantum fidelity 
\cite{Uhl,Jozsa,NC} between the initial and the evolved state emerges as a measure 
of their distinguishability.  Specifically, in Ref. \cite{GLM}, the QSLT is redefined 
as the time ${\tau}_{\epsilon}$ indicating how fast the fidelity between 
an initial state ${\hat \rho}(0)$ and the evolved one ${\hat \rho}(t)$ could reach 
a given value $\epsilon \in [\, 0,1)$.

Another major development was then achieved with the generalization of the QSLs 
to the evolution of open quantum systems \cite{Taddei,Campo,DL,Sun,UK,MTW,WuYu18}.  
For any open system initially prepared in a state ${\hat \rho}(0)$, and then left to evolve 
under Markovian or non-Markovian dynamics,  the rate of change of a carefully chosen 
measure of distinguishability between the initial state ${\hat \rho}(0)$ and  the evolved one 
${\hat \rho}(t)$  could be used to define a QSL.  According to this procedure, 
Taddei {\em et al}  \cite{Taddei} found an expression in terms of the quantum Fisher information, 
while del Campo {\em et al} \cite{Campo} established  upper bounds to the rate of change 
of the relative purity for both Markovian and non-Markovian systems.  A  treatment based 
on purity was recently proposed  in Ref.\cite{UK}.  Furthermore, Deffner and Lutz \cite{DL} 
derived geometric generalizations to open quantum systems of both limit times known 
for unitary evolutions, the Mandelstam-Tamm bound, as well as the Margolus-Levitin one. 
Note that the geometric QSLs previously proposed in Refs. \cite{Kok,Zwierz} are derived 
as upper bounds on the rate of change of a distance-type measure of distinguishability. 
General geometric QSLs are analyzed in Ref.\cite{Pires} using as measures of distinguishability  
a family of Riemannian metrics that are contractive under stochastic maps. Among them, 
geometric QSLs based on the quantum Fisher information metric (involving fidelity) \cite{GI} 
and the Wigner-Yanase information metric (involving affinity) \cite{LZ} were compared  
in Ref. \cite{Pires} for unitary dynamics and several examples of open-system evolutions. 
More recently,  some other QSLs were introduced, although they are not proper distance-type 
measures of quantum evolution. For instance, in Ref. \cite{MTW} one finds a definition based 
on a "quantumness" notion, while Ref.\cite{MDS} shows that quantum coherence plays 
an important role in setting the QSL.

It is worth emphasizing the importance of the feasibility of a QSL, defined as the easiness 
of evaluating the distance between the involved states \cite{Modi1,Modi2}. 
Finding tight and feasible QSLs for various evolutions became meanwhile an issue of interest 
in some areas of quantum information science \cite{EW,Campo1}, such as quantum computation 
\cite{SS}, quantum metrology \cite{GLM2011}, and quantum control \cite{DC,CD}, as we learn 
from the recent surveys \cite{Dod,Frey,DC,Modi2}. Feasibility could be achieved by avoiding  
the squared root of operators entering formulas based on fidelity or affinity,  which are used
in Refs.\cite{DL,Pires,MDS}  and employing instead  the Hilbert-Schmidt (HS) metric. 
In spite of all noticed inconsistencies of the HS metric \cite{Ozawa,Piani}, there is already 
a large amount of work on feasible QSLs in finite-dimensional systems using the HS metric 
\cite{Modi1,Modi2} or some Bures-metric relatives obtained by defining alternative formulas 
for the fidelity \cite{Sun,Modi2,EBF,WuYu20}. Especially interesting for continuous-variable 
systems, a feasible  treatment of QSLs based on the rate of change of the Wigner function 
was developed in Refs.\cite{Deffner2017,Campo2}.

In the present work we reconsider from the QSL-perspective the standard setting 
of a cavity mode of the radiation field interacting weakly with a thermal bosonic reservoir. 
The evolution of the one-mode field state is governed by a quantum optical master equation 
\cite{Louisell,WM,SG,PT93a,PT93b,PT2000a,PT2000b,Pav,PIT13}, which is
a continuous-variable prototype of the Lindblad-form equation \cite{Lindblad,BP,RH}. 
Whereas several geometric QSLs are employed so far to describe the evolution 
of finite-dimensional systems \cite{Campo,DL,Pires,Sun,Modi2},  we select two of them 
to be applied to the continuous-variable system specified above. These are based, 
respectively, on the quantum fidelity and the HS metric. We evaluate them by choosing
an initial pure state and making use of the quantum optical master equation, 
which is analytically solvable. 

A treatment based on the relative purity, which is developed in Ref. \cite{Campo} 
for a general Markovian evolution, allows one to get a time-independent QSL. 
This is determined only by the Lindblad form of the master equation and the initial state 
of the system. For any initial pure state, the relative purity reduces to the fidelity between 
the initial state ${\hat \rho}(0)$ and the evolved one ${\hat \rho}(t)$, 
\begin{equation}
{\cal F}(t):={\rm Tr}\left[ {\hat \rho}(0){\hat \rho}(t) \right],
\label{fidev}
\end{equation}
which is fairly termed {\em fidelity of evolution}.

In order to find time-dependent upper bounds to the speed of quantum evolution 
one usually applies a slightly different version of the method pointed out in 
Ref. \cite{Campo}. One starts from the rate of change for a convenient measure 
of distinguishability between the initial state and the evolved one. Doing so and then
employing the Cauchy-Schwarz inequality for HS operators, we recover lower bounds 
to the driving times, which are associated with the envisaged figures of merit: 
the fidelity of evolution and the HS evolution distance. To evaluate the latter, 
we need, in addition, the purity of the transient state:
\begin{equation} 
{\cal P}(t):= {{\rm Tr}\left\{ \left[ \hat \rho(t) \right]^2 \right\}} \leqq 1.
\label{purity}
\end{equation}
We evaluate both QSLTs by using the characteristic function (CF) of the evolved state 
${\hat \rho}(t)$, obtained by solving the quantum optical master equation. An interesting 
goal is to compare them for specified input states.

Explicit solutions are found and analyzed in some detail for two families of pure states 
which are currently used in quantum physics: the coherent and the Fock states. 
The former are the only pure states that are classical;  they are ubiquitous since the advent 
of the laser. The latter are the eigenstates of the energy operator of a free-field mode: 
${\hat H}= \hbar \omega \, {\hat a}^{\dag}{\hat a}$. Except for the vacuum state, 
they are highly non-classical. 

The paper is structured in the following way. Section II starts with a review of ideas 
and methods put forward in Refs.\cite{Campo,DL,Pires}. To the original treatment developed 
in Ref. \cite{Campo} we add an alternative fidelity-based QSL, which was first written
in Ref. \cite{DL} in connection with the Bures-angle measure and holds also for non-Markovian 
evolutions. Remarkably, the latter QSL turns out to coincide with that stemming from the rate 
of change of the HS distance between a pure initial state and the evolved one, as discussed 
earlier in Ref.\cite{Modi2}. Section III deals with the time-independent QSL obtained by using 
the fidelity of evolution, as initiated in Ref.\cite{Campo}. We derive here a general formula 
that holds for an arbitrary pure one-mode state decaying under the quantum optical master 
equation. Then we apply it to the coherent and number states. In Sec. IV, we review the CF method 
of describing the damping of a field mode governed by the quantum optical master equation.
Owing to Weyl's expansion formula, the main ingredients employed in this article 
are finally written as phase-space integrals. We perform two of them in Sec. V for coherent-state 
inputs and in Sec. VI for number-state ones, to get compact formulas for the fidelity of evolution 
and the HS evolution distance. We evaluate and analyze in Sec. V the time-dependent QSLs 
for thermalization of the coherent states. Some associate QSLTs are then compared.
In Sec. VI, we restrict ourselves to evaluate the time-dependent QSL for dissipation of one- and 
two-photon states. The pair of associate QSLTs is then examined. Our conclusions are summarized
in Sec. VII. The article includes three appendices. In Appendix A, which deals with coherent states 
other than vacuum, we justify the replacement of the Schr\"odinger-picture fidelity of evolution 
by its interaction-picture counterpart, made in Sec.V. Appendix B develops a straightforward 
method to perform the basic integral needed for describing the damping of the Fock states. 
In Appendix C we carefully analyze the decay of a one-photon state, revealing the non-monotonic 
behavior of the HS distance of evolution. 

\section{Quantum speed limits for Markovian open systems}

The Markovian dynamics of an open quantum system with the unperturbed Hamiltonian 
$\hat H$ is described by a Lindblad-type master equation for its density operator 
$\hat \rho(t)$ in the Schr\"odinger picture \cite{Lindblad}:
\begin{align}
& \frac{\partial \hat \rho(t)}{\partial t}={\cal L} \hat \rho(t):=-\frac{i}{\hbar} [\hat H, \hat \rho(t)]  
\notag  \\
& +\sum_{k} \left[ \hat L_k  \hat \rho(t) \hat L_k^{\dag}
-\frac{1}{2}\{ \hat L_k^{\dag} \hat L_k, \hat \rho(t) \} \right].
\label{L}
\end{align}
In Eq. (\ref{L}), ${\hat L}_k$ are the time-independent Lindblad operators and ${\cal L}$ 
is called the Liouville super-operator. The latter is the time-independent generator 
of a one-parameter family of quantum dynamical maps,
\begin{equation} 
{\cal V}(t):=\exp(t {\cal L}),  \qquad   (t \geqq 0),
\label{V} 
\end{equation}
displaying the semigroup property 
$${\cal V}(t_1+t_2)={\cal V}(t_1){\cal V}(t_2),   \quad (t_1 \geqq 0, \; t_2 \geqq 0),$$
and with ${\cal V}(0)$ being the identity map. It can be shown that any map  ${\cal V}(t)$, 
describing the state change of the open system over time $t$, represents a convex linear,
completely positive, and trace-preserving quantum operation \cite{BP}.

A bound to the speed of evolution from the initial state ${\hat \rho}(0)$ to an evolved one
${\hat \rho}(t)$ under the dynamical map (\ref{V})  can be defined in terms of the decay rate 
of an accepted measure of distinguishability between the states ${\hat \rho}(0)$ and 
${\hat \rho}(t)$. The original request for such a bound was that it should be determined only 
by the unperturbed Hamiltonian ${\hat H}$ and the Lindblad operators in Eq. (\ref{L}), 
as well as by the input state ${\hat \rho}(0)$ \cite{Campo}. Were this to be possible, then 
the associate lower bound to the actual driving time could lead to an inequality analogous 
to Eqs. (\ref{MT}) or (\ref{ML}), with no need of explicitly solving the master equation (\ref{L}). 

\subsection{Using the fidelity of evolution}

A convenient candidate to this procedure turned out to be the relative purity proposed
in Ref.\cite{Campo} and exploited also in Refs. \cite{DL,Modi1,WuYu18}:
\begin{equation} 
f(t):=\frac{{\rm Tr} [\hat \rho(0) \hat \rho(t)]}{{\rm Tr}\{ [\hat \rho(0)]^2 \} }.
\label{rp} 
\end{equation}

Recall that the fidelity of two arbitrary quantum states, $\hat \rho_1$ and $\hat \rho_2$, 
introduced by Uhlmann, has the intrinsic expression \cite{Uhl}:
\begin{equation}
{\cal  F}(\hat \rho_1,\hat \rho_2)=\left\{ {\rm Tr} \left[ \left( \sqrt{\hat \rho_1}\hat \rho_2\sqrt{\hat \rho_1}
\right)^{\frac{1}{2} }\right] \right\}^2. 
\label{fidelity}
\end{equation}
If at least one of the states $\hat \rho_1$ and $\hat \rho_2$ is pure, Uhlmann's fidelity 
simplifies to ${\cal  F}(\hat \rho_1,\hat \rho_2)={\rm Tr} (\hat \rho_1 \hat \rho_2). $

For a pure-state input to the master equation (\ref{L}), $\hat \rho(0)=|\Psi(0)\rangle \langle\Psi(0)|$,  
the relative purity (\ref{rp}) reduces thus to the fidelity of evolution (\ref{fidev}).
In Ref. \cite{Campo},  del Campo {\em et al.} found an upper bound for the rate of change 
of the fidelity of evolution (\ref{fidev}):
\begin{align}  
\dot{\cal F}(t)= {\rm Tr}[\hat \rho(0){\cal L} \hat \rho(t)]     
= {\rm Tr}[\hat \rho(t){\cal L}^{\dag} \hat \rho(0)].
\label{rateF}
\end{align} 
This is obtained by applying the Cauchy-Schwarz inequality for Hilbert-Schmidt (HS) operators, 
\begin{equation}
|{\rm Tr }({\hat A}^{\dag} \hat B)| \leqq || \hat A ||_2 || \hat B ||_2,
\label{CS}
\end{equation}
to the second form of Eq. (\ref{rateF}). The HS norm of such an operator $\hat A$ is defined as:
\begin{equation} 
|| \hat A ||_2:= \sqrt{{\rm Tr}({\hat A}^{\dag} \hat A )}. 
\label{HSnorm}
\end{equation}
Introducing the positive quantity
\begin{equation} 
v_{\cal F}(0):=\sqrt{{\rm Tr} \left\{ \left[ {\cal L} ^{\dag}\hat \rho(0)\right]^2 \right\}}
\label{vF0}
\end{equation}
and its non-static natural extension,
\begin{equation} 
v_{\cal F}(t):=v_{\cal F}(0)\sqrt{ {\cal P}(t) } \leqq v_{\cal F}(0),
\label{vFt}
\end{equation}
where ${\cal P}(t)$ is the purity (\ref{purity}), the Cauchy-Schwarz inequality 
for the second form of Eq. (\ref{rateF}) reads simply:
\begin{equation} 
\left| \dot{\cal F}(t) \right| \leqq v_{\cal F}(t).
\label{CS1}
\end{equation} 
Therefore, when choosing the fidelity (\ref{fidev}) as a feasible indicator of evolution,
both quantities (\ref{vF0}) and (\ref{vFt}) are speed upper bounds within the Markovian 
approach. The former, Eq. (\ref{vF0}), is looser, but is a time-independent QSL 
that can be evaluated without solving the master equation (\ref{L}). Indeed, this static 
speed limit is determined by the generator of the dynamical semigroup 
of the Markovian master equation (\ref{L}) and by the initial pure state of the system. 
That is why it was preferred in Ref.\cite{Campo}. The latter, Eq. (\ref{vFt}), albeit tighter,
has the drawback of being time-dependent, so that its evaluation requires the explicit 
solution of the master equation (\ref{L}).

It is convenient to employ the time-averaged speed of evolution
\begin{equation} 
\overline{v_{\cal F} }(t):=\frac{1}{t} \int_{0}^{t} dt^{\prime} \,v_{\cal F} (t^{\prime}) \leqq v_{\cal F}(0).
\label{QSL1}
\end{equation}
Integration of Eq. (\ref{CS1}) over the time interval $[0,t]$ provides the inequality
\begin{equation} 
1-{\cal F}(t) \leqq t \,\overline{v_{\cal F} }(t).
\label{1-F1}
\end{equation}
Accordingly, we introduce the conventional time
\begin{equation} 
{\tau}_{\cal F}(t):= \frac{1-{\cal F}(t) }{\overline{v_{\cal F} }(t) },
\label{tauF}
\end{equation}
as well as its minimum with respect to the variable $\overline{v_{\cal F} }(t)$ at fixed fidelity:
\begin{equation} 
{\tau}_{\cal F}^{\rm min}(t):= \frac{1-{\cal F}(t) }{v_{\cal F}(0) }.
\label{tauFmin}
\end{equation}
Both times (\ref{tauF}) and (\ref{tauFmin}) are lower bounds to the time of evolution 
from the pure state $\hat \rho(0)$ to the state $\hat \rho(t)$, \\ i. e., QSLTs:
\begin{equation}
t \geqq {\tau}_{\cal F}(t) \geqq {\tau}_{\cal F}^{\rm min}(t). 
\label{tautau}
\end{equation}

Let us mention that a similar treatment of Markovian dynamics, which is based 
on the purity (\ref{purity}) of the state as a decaying figure of merit, also avoids 
an explicit knowledge of the evolved state when deriving a QSLT. This is written 
in terms of the HS norms of the Lindblad operators $\hat L_k$ \cite{UK}. 
However,  this method is not applicable to continuous-variable systems, 
whose Lindblad operators are usually unbounded. 

For further insight,  we now choose to apply the Cauchy-Schwarz inequality 
in the first form of the rate of evolution (\ref{rateF}):  
\begin{align} 
\left| \dot{\cal F}(t) \right| \leqq  \sqrt{ {\rm Tr} \left\{ \left[ {\cal L}\hat \rho(t) \right]^{\dag}  
\left[ {\cal L} \hat \rho(t) \right] \right\} }
=\left| \left| \frac{\partial \hat \rho(t)}{\partial t} \right| \right|_2.
\label{CS2}
\end{align}
With the notation
\begin{equation}
{\tilde v}(t):=\left | \left | \frac{\partial \hat \rho(t)}{\partial t} \right| \right|_2,
\label{tv}
\end{equation}
Eq.\ (\ref{CS2}) takes the same form as Eq.\ (\ref{CS1}):
\begin{equation} 
\left| \dot{\cal F}(t) \right| \leqq {\tilde v}(t).
\label{rateF1}
\end{equation} 
By using the average of the bound (\ref{tv}) over the current time of evolution,
\begin{equation} 
\overline{ {\tilde v} }(t):=\frac{1}{t} \int_{0}^{t} dt^{\prime} \, \tilde{v} (t^{\prime}), 
\label{QSL2}
\end{equation}
we write the following inequality derived by integrating Eq.\ (\ref{QSL2}):
\begin{equation} 
1-{\cal F}(t) \leqq t \,\overline{ {\tilde v} }(t).
\label{1-F2}
\end{equation}
Equation\ (\ref{1-F2}) is an analogue of Eq.\ (\ref{1-F1}) and leads to an alternative QSLT:  
\begin{equation}
t \geqq \tilde{\tau}_{\cal F}(t):= \frac{1-{\cal F}(t)}{\overline{ \tilde{v} }(t) }.
\label{ttauF}
\end{equation}

Note that the QSL (\ref{QSL2}) can be evaluated only by making use of the explicit solution 
$\hat \rho(t)$ of the master equation\ (\ref{L}). The QSLT (\ref{ttauF}) is derived above
following the pure-state treatment of Deffner and Lutz in Ref.\cite{DL} and holds 
for non-Markovian dynamics as well.

\subsection{Using the Hilbert-Schmidt metric}

In quest of feasible QSLs, one could take inspiration from the considerable efforts and results 
on applications of distance-type measures in quantifying  various kinds of quantum correlations 
\cite{Adesso16}. Geometric measures based on the HS metric are known for a long time as 
working fairly well for quantification of non-Gaussianity \cite{Paris07,IPT13} or for the evaluation 
of quantum discord \cite{DVB}. However, we recall that the geometric measures of quantum 
correlations defined with the HS metric were found to display some inconveniences due to 
its non-contractivity under completely positive, trace-preserving maps \cite{Ozawa}. It will 
therefore be interesting to examine the behavior of a QSLT built with the HS metric \cite{Modi2}. 
We start by writing the HS distance between the initial pure state $\hat \rho (0)$ 
and the evolved one $\hat \rho (t)$:
\begin{equation}
{\cal G}(t):=|| \hat \rho (t)-\hat \rho (0)||_2 . 
\label{HS}
\end{equation}
The above {\em Hilbert-Schmidt distance of evolution} depends on both the fidelity 
of evolution ${\cal F}(t)$, Eq. (\ref{fidev}), and the purity ${\cal P}(t)$ of the evolved state, 
Eq. (\ref{purity}):
\begin{equation} 
{\cal G}(t)=\left[ 1+ {\cal P}(t)-2 {\cal F}(t) \right]^{\frac{1}{2} }.
\label{HS1}
\end{equation}
We apply the Cauchy-Schwarz inequality (\ref{CS}) to its rate of change,
\begin{align}
\left |\dot{\cal G}(t) \right | =2 \left |{\rm Tr} \left\{ \frac{\partial \hat \rho(t)}{\partial t}
\left[\hat \rho(t)-\hat \rho(0)\right)] \right\} \right | \frac{1}{2\, {\cal G}(t) } \leqq {\tilde v}(t):  
\notag 
\end{align}
\begin{equation} 
\left | \dot{\cal G}(t) \right | \leqq {\tilde v}(t).
\label{rateG} 
\end{equation}
Remarkably, as shown by Eqs. (\ref{rateF1}) and (\ref{rateG}), the rates of evolution 
of both the fidelity (\ref{fidev}) and the HS distance (\ref{HS}) share the upper bound 
${\tilde v}(t),$  Eq. (\ref{tv}). In addition, by integrating Eq. (\ref{rateG}) over the time 
interval $[0, t]$, one gets the inequality
\begin{equation} 
{\cal G}(t)  \leqq t \,\overline{ \tilde{v} }(t),
\label{G}
\end{equation}
providing the Hilbert-Schmidt QSLT:
\begin{equation} 
t \geqq \tilde{\tau}_{\cal G}(t):= \frac{ {\cal G}(t) }{\overline{ \tilde v }(t) }.
\label{ttauG} 
\end{equation}
We stress that the QSLT (\ref{ttauG}) was derived by Campaioli, Pollock, and Modi,
in Ref. \cite{Modi2}, regardless of the purity ${\cal P}(0)$ of the initial state, 
by employing an ingenious geometric method which holds for a finite-dimensional 
Hilbert space. 

The QSLTs (\ref{ttauF}) and (\ref{ttauG}) differ only by the figures of merit employed 
to distinguish the states $\hat \rho(0)$ and $\hat \rho(t)$.  This allows us to establish 
an inequality between these two QSLTs that is valid for any input pure state. 
Indeed, we apply once again the Cauchy-Schwarz inequality (\ref{CS}) to write:  
\begin{equation} 
{\rm Tr} [\hat \rho(0) \hat \rho(t)] \leqq ||\hat \rho (t)||_2:  \qquad   {\cal F}(t) \leqq  \sqrt{{\cal P}(t)}. 
\label{FP}
\end{equation}
Equation\ (\ref{FP}) implies the inequality 
\begin{equation}
{\cal G}(t)\geqq 1-{\cal F}(t)
\label{GF}
\end{equation}
and hence, 
\begin{equation}
\tilde{\tau}_{\cal G}(t)  \geqq  \tilde{\tau}_{\cal F}(t).
\label{ttauGF}
\end{equation}

Let us make clear what is usually meant by a tight QSL.  When employing contractive 
Riemannian metrics as measures of distinguishability between an initial pure state 
${\hat \rho}(0)$ and the evolved one ${\hat \rho}(t)$, one could identify the tightness 
of a QSL with the closeness of the given dynamical evolution to the corresponding 
geodesic \cite{Pires}.  In order to discuss the tightness of the QSLTs we deal with 
in the present work, we follow the simpler idea put forward in Refs.\cite{DL, Modi1, Modi2}
to compare how close they are to the actual time of evolution $t$.  For instance, 
according to Eq.(\ref{ttauGF}),  the QSLT $\tilde{\tau}_{\cal G}(t)$,  Eq.(\ref{ttauG}), 
expressed in terms of the HS metric, is tighter than the fidelity-based one, 
$\tilde{\tau}_{\cal F}(t)$, Eq.(\ref{ttauF}). 

Another distinctive property of the HS QSLT (\ref{ttauG}) is its robustness under 
composition \cite{Modi2}. Indeed, the addition of an uncorrelated ancillary system,
whose state is mixed and time-independent, changes the HS distance of evolution (\ref{HS}), 
multiplying it by the square root of the purity of the {\em ancilla} state \cite{Piani}.
Then the QSL (\ref{QSL2}) is multiplied by the same factor. As a consequence,
the HS QSLT $\tilde{\tau}_{\cal G}(t)$,  Eq.(\ref{ttauG}), remains unchanged when one modifies
the quantum system by adding or removing an uncorrelated {\em ancilla} in such a state.

We conclude this section with some few remarks. The treatment of del Campo {\em et al.} 
\cite{Campo} provides the time-independent QSL, Eq. (\ref{vF0}). Therefore, this is 
the unique bound that can be evaluated without explicitly solving the master equation\ (\ref{L}). 
We have derived three other QSLs that are time-dependent, Eq. (\ref{QSL1}) and  
Eq.(\ref{QSL2}) twice. This happens because, in view of Eqs. (\ref{rateF1}) and (\ref{rateG}), 
the last two QSLs coincide. Their evaluation requires the solution of the master equation (\ref{L}). 
However, the associate QSLTs,  Eqs. (\ref{tauF}),  (\ref{ttauF}), and (\ref{ttauG}), are tighter 
to the actual driving time $t$ than the QSLT (\ref{tauFmin}), which was introduced 
in Ref. \cite{Campo}. This last one may be used in a straightforward way, without solving 
the master equation (\ref{L}). For instance, one can readily evaluate the QSLT (\ref{tauFmin}) 
concerning the half-life $t_{1/2}$ of the input state, defined by the condition 
${\cal F}(t_{1/2})=\frac{1}{2}$.

\section{Quantum optical master equation:  the speed limit ${\bf v_{\cal F}(0) }$}

While a large amount of research has been done on the speed of evolution of finite-dimensional 
open quantum systems \cite{Frey,DC,Pires,Modi2}, analogous examples in the continuous-variable 
settings are scarce \cite{Deffner2017,Campo2}.  Two important Markovian master equations 
involving such a system were derived and frequently applied for a long time: 
the quantum optical master equation \cite{Louisell,WM,SG,PT93b,PT2000a,PT2000b} 
and the master equation for the quantum Brownian motion of a particle in high-temperature 
regime \cite{A71,CL,UZ,Zurek,Gallis}. Their solutions  were used to investigate the evolution 
of various properties of the open quantum system. For instance, in the case of quantum optical 
master equation, the alteration of coherence and squeezing under dissipation received 
considerable attention \cite{WM,SG,PT93b}, as well as the emergency of classicality for the field 
state \cite{Zurek,Gallis,Pav,PIT13}. In what follows, we apply the concepts and methods discussed 
in the preceding section to a continuous-variable system which is fundamental in quantum optics: 
a cavity mode of the quantum radiation field which interacts weakly with a thermal bosonic reservoir.

\subsection{Quantum optical master equation}

Here we choose to employ the master equation for a damped harmonic oscillator \cite{Louisell,BP}  
as a simplified version of quantum optical master equation \cite{BP}. The corresponding mode 
of the radiation field with the angular frequency $\omega$ and the amplitude operators
$\hat a$ and ${{\hat a}^{\dag}}$  is sustained by a cavity which stands for the environment. 
The effective interaction between the mode and the cavity consists in photon scattering 
by atoms into and out of the mode. These radiative processes produce a damping 
of the cavity mode characterized by the field-reservoir coupling constant $\gamma$. 
The mean number of photons $\bar{n}_{\rm R}$ in any mode of the reservoir having 
the same frequency  $\omega$ is given by the Bose-Einstein distribution function:
\begin{equation}
\bar{n}_{\rm R}=\left[ \exp{ \left( \frac{\hbar \omega}{k_{B}T} \right) } -1 \right]^{-1}.
\label{BE}
\end{equation}
Recall also the Hamiltonian $\hat H=\hbar \omega\, {\hat a}^{\dag} \hat a$ 
for the free evolution of the mode. Under the weak-coupling condition $\gamma \ll \omega$, 
one can describe the damping of the cavity mode by the following quantum optical master 
equationin the Schr\"odinger picture \cite{Louisell,BP}:
\begin{align} 
& \frac{\partial \hat \rho(t)}{\partial t}  = -i \omega  [\hat a^{\dag} \hat a,\, \hat {\rho}(t)]    \notag   \\ 
& +\gamma \bar{n}_R \left\{ \hat a^{\dag} \hat \rho(t) \hat a
-\frac{1}{2} [\hat a \hat a^{\dag} \hat \rho(t) +\hat {\rho}(t) \hat a \hat a^{\dag}] \right\}      \notag   \\ 
& +\gamma (\bar{n}_R+1) \left\{ \hat a \hat {\rho}(t) \hat  a^{\dag}
-\frac{1}{2}[\hat a^{\dag}\hat a \hat {\rho}(t)+\hat\rho(t) \hat a^{\dag} \hat a] \right\}.    
\label{QOME}
\end{align}
The master equation (\ref{QOME}) can be derived starting from the von Neumann equation
for the density operator of the field-reservoir system in the interaction picture, by applying 
three distinct approximations, namely, the Born, Markov, and rotating wave approximations \cite{BP}.   
The Markov approximation is valid provided that a strong inequality between the relaxation times 
of the field mode, ${\tau}_F$, and the reservoir, ${\tau}_R$, is fulfilled: ${\tau}_F \gg {\tau}_R$.
The former has the order ${\tau}_F \approx {\gamma}^{-1}$, where $\gamma$ is essentially
a typical value for the rates of the electric dipole transitions in atoms, i. e., $10^7 \,{\rm s}^{-1}$
to $10^9\,{\rm s}^{-1}$. On the other hand, the latter is of same order as the vacuum correlation time
of the bosonic reservoir, which is given by the inverse of a typical transition frequency 
in the optical domain: ${\tau}_R \approx {\omega}^{-1} \approx 10^{-15} \,{\rm s}$.
To sum up, the condition of validity for the Markov approximation, ${\tau}_F \gg {\tau}_R$, 
coincides with that for the Born approximation, namely,  the weak-coupling condition 
$\gamma \ll \omega$. This is largely fulfilled in the optical domain, where
$\gamma / \omega \approx 10^{-6}$. Finally, the rotating wave approximation is justified 
as well, since the terms containing factors $\exp[i({\omega}^{\prime}-\omega)t]$ with
${\omega}^{\prime} \neq \omega$ are rapidly oscillating on the time scale
${\tau}_F \approx {\gamma}^{-1}$ of the mode damping and can therefore be neglected.

The quantum optical master equation (\ref{QOME}) has the Lindblad form and therefore
it preserves the positivity of the density operator.  As a consequence, the Robertson-Schr\"odinger 
uncertainty relation holds at any time \cite{SS87}. By comparison of Eqs. (\ref{QOME}) and (\ref{L}), 
we identify two unbounded Lindblad operators:
$$\hat L_1=\sqrt{\gamma \bar n_R}\;  \hat a^{\dag},    \;\;\;
\hat L_2=\sqrt{\gamma ( \bar n_R+1)}\;  \hat a.$$

\subsection{The quantum speed limit ${\bf v_{\cal F}(0) }$ }
  
We consider an arbitrary pure input one-mode state $\hat \rho(0)=|\Psi(0)\rangle \langle\Psi(0)|$,  
After a routine calculation we get the square of the QSL $v_{\cal F}(0)$,  Eq. (\ref{vF0}):
\begin{align}
& \left[ v_{\cal F}(0) \right]^{2} =\frac{2}{ {\hbar}^2 }
\left[ 1+\left( \bar n_R+\frac{1}{2}\right)^2 \frac{\gamma^2} {\omega^2}\right] 
[ (\Delta E)_0 ]^2         \notag  \\
& +4\gamma \omega \, \Im \left( \langle \hat a \rangle_0 \langle {\hat a}^{\dag} \hat a  {\hat a}^{\dag}
\rangle_0 \right)           \notag  \\ 
& +{\gamma}^2 \left\{  2\left[  3\bar n_R (\bar n_R+1)+1 \right] \langle{ \hat a}^{\dag} {\hat a}\rangle _0
\, ( \langle {\hat a}^{\dag} {\hat a}\rangle_0 +1)   \right.  \nonumber   \\
&  \left. + 2\bar n_R (\bar n_R+1)\left( \left| \langle {\hat a}^2 \rangle _0 \right|^2 
+1 \right) +1    \right.     \notag  \\ 
&  \left. -2(2{\bar n}_R+1)^2 \Re( \langle \hat a \rangle_0
\langle {\hat a}^{\dag} \hat a {\hat a}^{\dag} \rangle_0 ) \right\}.
\label{v(0)}
\end{align}
In Eq. (\ref{v(0)}) as well as in subsequent ones, the subscript $0$ of an expectation value means 
that it is taken at $t=0$. 

The no-damping limit of Eq. (\ref{v(0)}), $\gamma =0$, specifies the QSL
(\ref{vF0}) for the free evolution of the field mode:
\begin{equation} 
v_{\cal F}(0)=\sqrt{2}\,\frac{(\Delta E)_0}{\hbar}.
\label{free}
\end{equation} 
The corresponding QSLT (\ref{tauFmin}) reads:
\begin{equation} 
{\tau}_{\cal F}^{\rm min}(t)=[1-{\cal F}(t)] \,\frac{\sqrt{2} }{2} \frac{\hbar}{(\Delta E)_0}.
\label{QSLTfree}
\end{equation} 
Recall that the passage time $t_{\bot}$ under a unitary pure-state evolution is defined 
as the minimum time required for the evolving state $|\Psi(t)\rangle \langle\Psi(t)|$ to become 
othogonal to its initial value: ${\cal F}(t_{\bot})=0$. As a particular case of the general formula 
derived in Ref. \cite{Campo},  the passage time inferred from Eq. (\ref{QSLTfree}) has a bound 
considerably looser than the Mandelstam-Tamm bound (\ref{MT}) for an isolated system:
\begin{equation} 
{\tau}_{\cal F}^{\rm min}(t_{\bot})=\frac{\sqrt{2} }{2} \frac{\hbar}{(\Delta E)_0}
=\frac{\sqrt{2} }{\pi} {\tau}^{\rm (MT)}_{\bot}.
\label{QSLTsfree}
\end{equation} 
For a zero-temperature reservoir $(\bar n_R=0)$, the squared QSL (\ref{v(0)}) simplifies to:
\begin{align}
& \left[ v_{\cal F}(0) \right]^{2} =\frac{2}{ {\hbar}^2 }
\left( 1+ \frac{1}{4} \frac{\gamma^2} {\omega^2} \right) [ (\Delta E)_0 ]^2     \notag  \\
& +4\gamma \omega \, \Im \left( \langle \hat a \rangle_0 \langle {\hat a}^{\dag} \hat a  {\hat a}^{\dag}
\rangle_0 \right)       \notag  \\
& +{\gamma}^2 \left[ 2\langle {\hat a}^{\dag} {\hat a}\rangle_0 \,
( \langle {\hat a}^{\dag} {\hat a}\rangle_0 +1) +1   \right.       \notag   \\ 
&   \left. -2\, \Re( \langle \hat a \rangle_0 \langle {\hat a}^{\dag} \hat a {\hat a}^{\dag} \rangle_0 ) \right].
\label{solT=0}
\end{align}
We apply the general formula (\ref{v(0)}) to a couple of families of pure states that are 
ubiquitously employed in quantum optics.  The first one consists of the coherent states, 
which are usually defined as the eigenstates of the photon annihilation operator \cite{Glauber}:
$$\hat a |\alpha \rangle =\alpha |\alpha \rangle,  \qquad  (\alpha \in {\mathbb C}).$$  
The coherent states
are Gaussian and classical.
Moreover, they are the only 
classical pure states \cite{Cahill}. For any coherent state $|\alpha \rangle \langle \alpha|$, we employ 
the expectation value
$$\langle \alpha | (\hat a^{\dag})^l{ \hat a}^m |\alpha \rangle= (\alpha^*)^l{\alpha}^m$$
to evaluate the squared QSL (\ref{v(0)}):
\begin{align}
& \left[ v_{\cal F}(0) \right]^{2} 
=2 \left( \omega^2 +\frac{1}{4} {\gamma}^2 \right) |\alpha |^2        \notag  \\
& +\gamma^2 \left[ 2\bar n_R (\bar n_R +1) +1 \right].            
\label{v(0)coh}
\end{align}

When starting at $t=0$ from a coherent state, a free field mode evolves in a time-dependent 
coherent state: 
$$\exp{ \left[ -i \omega t (\hat a^{\dag} \hat a)  \right] } |\alpha\rangle = |\alpha(t)\rangle \;\; 
{\rm  with} \;\;  \alpha(t)=\alpha \exp(-i \omega t).$$
The corresponding formulas
\begin{equation} 
E:=\langle {\hat H}\rangle =|\alpha|^2 \hbar \omega,  \quad   E_g=0, 
\quad    \Delta E =|\alpha| \, \hbar \omega
\label{E}
\end{equation} 
provide the following unified bound (\ref{LT}):
\begin{equation}
{\tau}_{\bot}=\frac{\pi}{\omega \, |\alpha| } \left( |\alpha|+1 -|\, |\alpha|-1| \right)^{-1}.   
\label{LTcoh}
\end{equation}
This is a strictly decreasing, continuous function of $|\alpha|$ defined on ${\mathbb R}_{+}$.
If $\alpha \neq 0$, the bound (\ref{LTcoh}) is finite and unattainable, since the fidelity of evolution
$${\cal F}({\tau}_{\bot})=| \langle \alpha| \alpha({\tau}_{\bot}) \rangle|^2
=\exp{\left( -2|\alpha |^2 \right) }.$$
does not vanish. Nevertheless, it becomes tighter in the high-intensity regime $|\alpha| \gg 1$, 
for which we mention the vanishing  limits:
$$\lim_{|\alpha| \to \infty} ({\tau}_{\bot})=0,  \qquad   
\lim_{|\alpha| \to \infty}[{\cal F}({\tau}_{\bot})]=0.$$

On the other hand, the no-damping limit $\gamma =0$ of the square root 
of the non-negative quantity (\ref{v(0)coh}) is the QSL (\ref{free}) for the free evolution 
of the field mode in a coherent state:
\begin{equation} 
v_{\cal F}(0)=\sqrt{2}\, |\alpha| \, \omega.
\label{freecoh}
\end{equation} 
The associate QSTL (\ref{QSLTsfree}) for the passage time reads:
\begin{equation} 
{\tau}_{\cal F}^{\rm min}(t_{\bot})=\frac{\sqrt{2} }{2} \frac{1}{ |\alpha| \, \omega}
=\frac{\sqrt{2} }{\pi} {\tau}^{\rm (MT)}_{\bot}.
\label{QSLTsfreecoh}
\end{equation} 

The second class of states we are dealing with is that of the Fock states, 
i. e., the eigenstates of the photon-number operator:
$$(\hat a^{\dag} \hat a) |M \rangle =M |M \rangle,   \qquad     (M=0, 1, 2, 3, ...) .$$
Accordingly, their characteristic property is
\begin{equation} 
(\Delta E )_0 = 0.
\label{DE=0}
\end{equation} 
Any excited Fock state $|M \rangle \langle M|, \; (M>0),\;$ is neither Gaussian, 
nor classical. By use of the expectation value
$$\langle M| (\hat a^{\dag})^l{ \hat a}^m|M\rangle=\frac{M!}{(M-m)!} \, \delta_{lm},$$
we get the squared QSL (\ref{v(0)}) of an arbitrary Fock state:
\begin{align}
& \left[ v_{\cal F}(M;0) \right]^2
=\gamma^2 \left\{ 2\left[ 3\bar n_R (\bar n_R+1)+1 \right] M(M +1)  \right.   \notag   \\
& \left. + 2\bar n_R(\bar n_R+1)+1 \right\}.
\label{v(M;0)}
\end{align}

Let us now consider the free evolution of the field mode $(\gamma =0)$ 
in an arbitrary Fock state. Then, in view of the stationarity condition (\ref{DE=0}), 
the QSL (\ref{free}) vanishes: $v_{\cal F}(M;0)=0$. Consequently, the QSTL (\ref{QSLTsfree}) 
of the passage time becomes infinite:
\begin{equation} 
{\tau}_{\cal F}^{\rm min}(M; t_{\bot})=\frac{\sqrt{2} }{\pi} {\tau}^{\rm (MT)}_{\bot M }=\infty.
\label{QSLTsfreeM}
\end{equation} 
At the same time, the Margolus-Levitin bound (\ref{ML}) remains finite
for any excited Fock state, 
$$ {\tau}^{\rm (ML)}_{\bot M}=\frac{\pi}{2} \, \frac{1}{M \omega},  \qquad  (M>0), $$
so that, by virtue of Eq. (\ref{LT}), it is physically irrelevant.

For a non-zero coupling $\gamma >0$, the most striking difference between the evolutions 
of the above two classes of states in the optical domain is due to the dominant 
frequency-dependent term in Eq. (\ref{v(0)coh}). Indeed, the influence of the thermal reservoir 
on the fast evolution of an initial coherent state is extremely small. On the contrary, 
Eq. (\ref{v(M;0)}) shows that the field-reservoir interaction fully determines the slow decay 
of an initially stationary state.

Remark also that the squared QSLs (\ref{v(0)coh}) and (\ref{v(M;0)}) are increased 
by the thermal noise $\bar n_R$ of the reservoir, Eq. (\ref{BE}), as well as by their input 
mean number of photons $|\alpha|^2$ and $M$, respectively. In both equations, 
the last two terms proportional to ${\gamma}^2$ coincide and describe an influence 
of the environment which is independent of $|\alpha|^2$ or $M$. Moreover, in each of them 
there is a ${\gamma}^2$ term that is proportional to $|\alpha|^2$ and $M$, respectively. 
Beyond these similarities, Eq. (\ref{v(M;0)}) includes additional ${\gamma}^2$ terms 
in comparison with Eq. (\ref{v(0)coh}), stemming from the field-reservoir interaction. 
Therefore, this interaction depends significantly on the input state of the field mode.  

The relevance of the input states we have chosen to deal with in this paper is nicely 
illustrated by the non-classicality properties of the photon-added coherent states.  
In Ref. \cite{PT2020}, one evaluates the maximal value $Q_p^{\rm max}(\alpha, {\alpha}^{\ast} )$ 
of the Husimi $Q$ function of a coherent state $|\alpha \rangle \langle \alpha|$ with $p$ 
added photons. Its HS degree of non-classicality, equal to  
$1-\pi \, Q_p^{\rm max}(\alpha, {\alpha}^{\ast} )$,
is shown to decrease with the coherent  intensity $|\alpha|^2$ and to increase 
with the photon number $p$. It follows that in the class of photon-added coherent states, 
the coherent ones $(p=0)$ display the least non-classicality, while the excited Fock states 
$(\alpha=0, \, p>0)$ are the most non-classical. In particular, it is proven that the HS degree 
of non-classicality of any Fock state increases with its number of photons. 
On the same lines, in Ref. \cite{ExtremalQSs}, both the coherent and Fock states 
are the simplest examples of what are now termed {\em extremal quantum states}. 
The sense of this concept is that, for instance, the photon-added coherent states 
appear to be intermediate between the coherent ones, which possess the least 
quantumness, and the excited Fock states, which exhibit the most quantumness.

\section{Damping of a field mode}

In order to apply the general ideas outlined in Sec. II,  we find it suitable to exploit 
some analytic solutions of the quantum optical master equation (\ref{QOME}). 
A straightforward way to proceed is to employ the normally-ordered characteristic 
function (NCF) of the evolving one-mode state $\hat \rho(t)$ \cite{CG}:
\begin{align}
\chi^{(N)}(\lambda, {\lambda}^{\ast}, t):= {\rm Tr}\left[ \hat \rho(t) \exp(\lambda \hat a^{\dag} )
\exp(-{\lambda}^{\ast} \hat a) \right].
\label{NCF}
\end{align}
The master equation (\ref{QOME}) can be converted into a linear first-order partial 
differential equation for the NCF (\ref{NCF}), as shown in Refs. \cite{Louisell, PT93b}:
\begin{align}
& \frac{\partial \chi^{(N)}}{\partial t} =-\gamma \bar{n}_{\rm R} |\lambda|^2 \chi^{(N)}
-\left( \frac{\gamma}{2}-i\omega \right) \lambda \frac{\partial \chi^{(N)}}{\partial \lambda}     
\notag   \\
& -\left( \frac{\gamma}{2}+i\omega \right) {\lambda}^{\ast}
\frac{\partial \chi^{(N)} }{\partial {\lambda}^{\ast} }.
\label{eqNCF} 
\end{align}
By using the method of the characteristic curves \cite{WU}, we get the explicit solution 
of Eq. (\ref{eqNCF}) as the following product expressed in terms of its given initial form 
$\chi^{(N)}(\lambda, {\lambda}^{\ast}, 0)$ :
\begin{align} 
\chi^{(N)}(\lambda, {\lambda}^{\ast}, t)= \chi^{(N)}(\lambda(t), {\lambda}^{\ast}(t), 0) 
\exp{ \left[ -\bar{n}_T(t) |\lambda|^2 \right] }              \notag  \\
\label{NCFt}
\end{align}
In Eq. (\ref{NCFt}), we have introduced the time-dependent parameter  
\begin{equation} 
\lambda(t):=\lambda \exp{ \left[ -\left( \frac{\gamma}{2}-i\omega \right) t \right] },
\label{lambda}
\end{equation} 
while the occurring exponential factor is precisely the NCF of a thermal state (TS),
\begin{equation}
{\chi}_{\rm T}^{(N)}(\lambda, {\lambda}^{\ast}, t) = \exp{ \left[ -\bar{n}_{\rm T}(t) |\lambda|^2 \right] }, 
\label{TS}
\end{equation} 
with the mean photon occupancy at time $t$: 
\begin{equation}
\bar{n}_{\rm T}(t):=\bar{n}_{\rm R}\, [1-\exp(-\gamma t) ].
\label{nT}
\end{equation} 

Recall the Weyl expansion of an evolving one-mode density operator \cite{Weyl}:
\begin{equation} 
\hat \rho(t)=\frac{1}{\pi}\int{\rm d}^{2}{\lambda}\; \chi(\lambda, {\lambda}^{\ast}, t) 
\hat D(-\lambda, -{\lambda}^{\ast}),
\label{W}
\end{equation} 
where 
\begin{align}
& \hat D(\lambda, {\lambda}^{\ast}):=\exp(\lambda \hat a^{\dag}-\lambda^{\ast} \hat a)     
\notag   \\
& = \exp{\left( -\frac{1}{2}|\lambda|^2 \right) }
\exp(\lambda \hat a^{\dag}) \exp(-{\lambda}^{\ast} \hat a)
\label{D}
\end{align}
is a Weyl displacement operator \cite{Glauber},  whose expectation value,
\begin{align}
& \chi(\lambda, {\lambda}^{\ast}, t):= {\rm Tr}\left[ \hat \rho(t) \hat D(\lambda, {\lambda}^{\ast}) \right]
\notag  \\
& = \exp{\left( -\frac{1}{2}|\lambda|^2 \right) }\chi^{(N)}(\lambda, {\lambda}^{\ast}, t),
\label{CF}
\end{align} 
is the CF of the state $\hat \rho(t)$. Here and subsequently, we denote an area element 
in the plane of the complex integration variable $\lambda$ 
by ${\rm d}^{2}{\lambda}:={\rm d}\Re({\lambda}) \, {\rm d}\Im ({\lambda})$. We also mention
the HS orthonormalization relation of the unitary displacement operators: 
\begin{equation}
{\rm Tr} \left[ {\hat D}^{\dag}(\lambda, {\lambda}^{\ast}) {\hat D}(\mu, {\mu}^{\ast}) \right]
= \pi  \delta^{(2)}(\lambda-\mu),
\label{TrDD}
\end{equation} 
with $\, \delta^{(2)}(\lambda-\mu):= \delta[\Re(\lambda-\mu) ] \, \delta[\Im(\lambda-\mu) ].$

The simplest consequences of the multiplication law (\ref{NCFt}) of NCFs 
are two addition rules, which reveal its significance:
\begin{align} 
& \langle \hat a  \rangle(t) = \langle \hat a  \rangle_0 \exp{ \left[ -\left( \frac{\gamma}{2}
+i\omega \right) t \right] } +\langle \hat a  \rangle_{\rm T}(t),               \notag  \\
& {\rm  where}  \quad   \langle \hat a  \rangle_{\rm T}(t) =0, 
\label{a+a}
\end{align} 
\begin{equation}
\langle {\hat a}^{\dag} \hat a  \rangle(t) = \langle {\hat a}^{\dag} \hat a  \rangle_0 \exp( -\gamma t ) 
+ \bar{n}_{\rm R}\left[ 1-\exp(-\gamma t) \right].
\label{n+n}
\end{equation} 
Indeed, the above couple of equations shows that the NCF (\ref{NCFt}) describes 
the linear superposition of two field modes which have the same defining features 
(frequency, direction of propagation, and polarization), but are in distinct evolving states: 
the former is an attenuated mode whose initial state ${\hat \rho}(0)$ may be chosen at will, 
while the latter is the thermal mode with the time-increasing mean photon occupancy 
$\bar{n}_{\rm T}(t)$,  Eq. (\ref{nT}).  According to the addition rule (\ref{n+n}), $\bar{n}_{\rm T}(t)$ 
is the average number of photons transferred in the time interval $[0, t]$ from the thermal 
reservoir into the cavity mode.

On the other hand, at sufficiently large times $(\gamma t \gg 1)$, the transient one-mode state 
${\hat \rho}(t)$ of the radiation field with the NCF (\ref{NCFt}), tends to a steady-state regime
by reaching the TS $\, {\hat \rho}_{\rm T}( {\bar n}_{\rm R} )$, whose mean photon occupancy 
(\ref{BE}) is imposed by the reservoir. Stated concisely, the decay of the field mode 
governed by the quantum optical master equation (\ref{QOME}) is a {\it thermalization} process, 
i. e., an irreversible evolution towards thermal equilibrium.
 
In order to compare the feasibility and tightness of the QSLTs discussed in Sec. II,
we have to evaluate some necessary ingredients: the fidelity of evolution 
${\cal F}(t)$, Eq. (\ref{fidev}), the purity ${\cal P}(t)$ of the evolved state, Eq. (\ref{purity}), 
and the bound of the speed of evolution ${\tilde v}(t)$, Eq. (\ref{tv}). By applying the Weyl 
expansion formula (\ref{W}) in conjunction with the orthonormalization property (\ref{TrDD}), 
we get the following general expressions in terms of the CF 
$\chi(\lambda, {\lambda}^{\ast}, t)$, Eq. (\ref{CF}): 
\begin{equation}
{\cal F}(t)=\frac{1}{\pi}\int{\rm d}^{2}{\lambda}\; \chi^*(\lambda, {\lambda}^{\ast}, 0)
\chi(\lambda, {\lambda}^{\ast}, t), 
\label{fidev1}
\end{equation}
\begin{equation}
{\cal P}(t)=\frac{1}{\pi}\int{\rm d}^{2}{\lambda}\; |\chi(\lambda,  {\lambda}^{\ast}, t)|^2,
\label{pur}
\end{equation}
and
\begin{equation}
[{\tilde v}(t)]^2=\frac{1}{\pi}  \int {\rm d}^{2}{\lambda}
\left| \frac{\partial \chi(\lambda,  {\lambda}^{\ast}, t) }{\partial t} \right|^2.
\label{v^2}
\end{equation}
It is known for a long time that purity is important in its own right when analyzing 
from a quantum-optical perspective the decay of a single mode of the radiation field 
in a dissipative environment \cite{PT2000a,PT2000b}.

Note that the Wigner quasiprobability distribution $W(\beta, {\beta}^{\ast}, t)$ 
is the Fourier transform of the CF $\chi(\lambda, {\lambda}^{\ast}, t),$ \cite{CG}:
\begin{align}
W(\beta, {\beta}^{\ast}, t) =\frac{1}{\pi}  \int {\rm d}^{2}{\lambda} 
\exp{ \left(\beta {\lambda}^{\ast}- {\beta}^{\ast}\lambda \right)}  \,             
\chi (\lambda, {\lambda}^{\ast}, t).
\notag  
\end{align}
Owing to Parseval's formula, the integral representations  (\ref{fidev1})-(\ref{v^2}) may be 
paralleled with similar ones, where the CF $\chi$ is replaced by the Wigner function $W$.
For instance, Eq. (\ref{v^2}) is equivalent to the formula:
\begin{eqnarray}
[{\tilde v}(t)]^2=\frac{1}{\pi}  \int {\rm d}^{2}{\beta}\left| \frac{\partial W (\beta, {\beta}^{\ast} ,t)}
{\partial t} \right|^2.
\label{v^2W}
\end{eqnarray}
In Ref. \cite{Deffner2017}, the r.h.s of Eq.(\ref{v^2W}) was shown to represent the rate 
of the Wasserstein-2-distance between time-dependent Wigner distributions. 
Equations (\ref{v^2}) and (\ref{v^2W}) display the explicit equality of the two QSLs defined 
in Ref. \cite{Deffner2017} and provide an explanation for Fig.1 therein.

To conclude this section, Eqs. (\ref{fidev1})-(\ref{v^2}) are general formulas obtained
by employing the continuous-variable  Weyl expansion (\ref{W}).  They are useful 
when one applies the straightforward CF method to solve any master equation that rules 
the evolution of a continuous-variable system. We could readily prove the equivalence 
between the CF method and the Wigner-function approach developed in Ref.\cite{Deffner2017}. 
In the next two sections we will take advantage of the benefits of describing
some interesting states of the damped radiation field by their CFs.

\section{Thermalization of a coherent state}

\subsection{Evolved state}

It is well known that the quantum optical master equation (\ref{QOME}) preserves
the Gaussian character as well as the classicality of the initial state \cite{PT93b}. 
An input coherent state ${\hat \rho}(0)=|\alpha \rangle \langle \alpha|$ is Gaussian, 
since its NCF is the exponential $\chi^{(N)}(\lambda, {\lambda}^{\ast}, 0)
=\exp{\left( {\alpha}^{\ast} \lambda-\alpha{ \lambda}^{\ast} \right)},$
and lies at the limit of classicality, since its Glauber-Sudarshan $P$ representation
is a Dirac $\delta$ distribution: $P(\beta, {\beta}^{\ast}, 0)= \delta^{(2)}(\beta-\alpha)$.

The transient NCF (\ref{NCFt}) reads: 
\begin{align}
\chi^{(N)} (\lambda, {\lambda}^{\ast}, t) & =\exp\left[ -{\bar n}_{\rm T}(t) |\lambda|^2  
+{\alpha}^{\ast}(t) \lambda -\alpha (t) {\lambda}^{\ast} \right] 
\label{NCFtcoh}
\end{align} 
where 
\begin{equation}
\alpha(t):=\alpha \exp{\left[ -\left( \frac{\gamma}{2}+i \omega \right) t \right] },
\label{alpha(t)}
\end{equation} 
in accordance with  Eq. (\ref{a+a}). It describes an evolving displaced thermal state (DTS): 
\begin{equation}
{\hat \rho}_{\, \rm DT}(t)= \hat D[\alpha(t), {\alpha}^{\ast}(t)] \, {\hat \rho}_{\rm T} [{\bar n}_{\rm T} (t)]
\, \hat D^{\dag} [\alpha(t), {\alpha}^{\ast}(t) ].
\label{DTS}
\end{equation} 
The one-mode DTSs are special Gaussian states, whose statistical properties 
are systematically studied in Ref. \cite{PT93a}. The expected extremity limits 
of the evolved state (\ref{DTS}), 
$$\lim_{t \to 0} \left[ {\hat \rho}_{\, \rm DT}(t)  \right] =|\alpha \rangle \langle \alpha|,    \qquad
\lim_{t \to \infty} \left[ {\hat \rho}_{\, \rm DT}(t)  \right] = {\hat \rho}_{\rm T}({\bar n}_{\rm R}),$$
are readily checked. 

Being the Fourier transform of the Gaussian function 
${\pi}^{-1} \chi^{(N)} (\lambda, {\lambda}^{\ast}, t)$ \cite{CG}, the Glauber-Sudarshan 
quasiprobability distribution is also a regular Gaussian function for $t>0$:
$$P(\beta, {\beta}^{\ast}, t)=\frac{1}{\pi {\bar n}_{\rm T}(t) } \exp \left[ -\frac{ |\beta -\alpha(t)|^2 }
{ {\bar n}_{\rm T}(t) } \right] ,    \quad    (t>0),$$ 
with the appropriate limit  $\delta^{(2)}(\beta-\alpha)$ at $t=0$. 

We mention two extreme situations that are insightful:
\begin{enumerate}
\item{ {\em Dissipation of a coherent state} $( \alpha \neq 0, \, {\bar n}_{\rm R}=0 )$.  \\
The evolving state (\ref{DTS}) becomes the coherent state $|\alpha(t) \rangle \langle \alpha(t)|$
with the attenuated oscillating amplitude (\ref{alpha(t)}).}
\item{ {\em Thermalization of the vacuum state} $( \alpha=0, \, {\bar n}_{\rm R}>0 )$. \\
If ${\hat \rho}(0)=|0 \rangle \langle 0|$, then the transient state (\ref{DTS}) reduces 
to the evolving TS $\, {\hat \rho}_{\rm T} [{\bar n}_{\rm T} (t)].$ }
\end{enumerate}

\subsection{Main ingredients}

At this point, we find it convenient to introduce the time-decreasing variable
\begin{equation}
\eta:=\exp(-\gamma t),      \qquad       \eta \in (0,1].
\label{eta}
\end{equation}
In order to evaluate the integrals (\ref{fidev1})-(\ref{v^2}) for the evolving DTS (\ref{DTS}), 
we make use of Eqs. (\ref{CF}) and (\ref{NCFtcoh}), finding the exact explicit formulas:
\begin{align}
{\cal F}(t)=\frac{1}{ 1+{\bar n}_{\rm T}(t) } \exp\left[ -|\alpha|^2 \,
\frac{ 1-2\sqrt{\eta} \cos(\omega t) +\eta }{ 1+{\bar n}_{\rm T}(t) } \right],
\label{Fcoh}
\end{align} 
\begin{equation}
{\cal P}(t)=\frac{1}{ 1+2{\bar n}_{\rm T}(t) } \, ,
\label{Pcoh}
\end{equation} 
\begin{align}
& \left[ {\tilde v}(t) \right]^2= 2\left[ {\omega}^2 +\left( \frac{1}{2}\gamma \right)^2 \right] 
|\alpha|^2  \frac{\eta} {\left[ 1+2{\bar n}_{\rm T}(t) \right]^2}     \notag  \\    
& +2\left( \gamma {\bar n}_{\rm R} \right)^2
\frac{ {\eta}^2 } {\left[ 1+2{\bar n}_{\rm T}(t) \right]^3} \, .
\label{vcoh^2}
\end{align} 
An analysis of the above functions will suggest us how to exploit them in an efficient way.

\subsubsection{Fidelity of evolution}

The fidelity of evolution (\ref{Fcoh}) is equal, up to a factor $\pi$, to the Husimi $Q$ function 
of the transient DTS (\ref{DTS}): 
\begin{equation}
{\cal F}(t)=\langle \alpha |{\hat \rho}_{\, \rm DT}(t)| \alpha \rangle
=:\pi \, Q_{\, \rm DT}( \alpha, {\alpha}^{\ast}, t).
\label{FcohSP}
\end{equation} 
Its value at thermal equilibrium,
$$\lim_{t \to \infty}[{\cal  F}(t)]=\langle \alpha |{\hat \rho}_{\rm T}( {\bar n}_{\rm R} )|\alpha \rangle
=:\pi Q_{\rm T}( {\bar n}_{\rm R}; \alpha, {\alpha}^{\ast} ),$$
reads:
\begin{equation}
\lim_{t \to \infty} \left[ {\cal F}(t) \right]=\frac{1}{ 1+{\bar n}_{\rm R} } 
\exp\left( -\frac{ |\alpha |^2}{1+ {\bar n}_{\rm R} } \right).
\label{FcohIO}
\end{equation}

If $\alpha \neq 0$, the fidelity of damping (\ref{Fcoh}) suffers from the drawback that, 
under the weak-coupling condition $\gamma \ll \omega$, and at the scale 
of the field relaxation time ${\tau}_F \approx {\gamma}^{-1},$  it oscillates very rapidly 
due to the presence of the function $\cos(\omega t)$ at the exponent. In the present case, 
this is a major inconvenience concerning the figures of merit occurring in the QSLTs 
(\ref{tauF}), (\ref{ttauF}), and (\ref{ttauG}). In principle, one can remedy it by replacing 
the oscillating function $\cos(\omega t)$ either with one of its convenient values 
or with a significant average connected to it. In Appendix A, we point out the appropriate 
approach of this kind. The conclusion is that the fidelity of evolution in the interaction picture 
is the only suitable one. This means that instead of the oscillating function (\ref{Fcoh}), 
we will use exclusively the smooth fidelity
\begin{align} 
{\cal F}^{(1)}(t)=\frac{1}{ 1+{\bar n}_{\rm T}(t) } 
\exp \left[ -|\alpha|^2 \, \frac{ \left( 1-\sqrt{\eta} \right)^2 }{ 1+{\bar n}_{\rm T}(t) } \right].      
\label{F1}
\end{align} 

Let us mention the short-time approximation of the function ${\cal F}^{(1)}(t)$, Eq. (\ref{F1}):
\begin{align}
& {\cal F}^{(1)}(t)=1-{\bar n}_{\rm R} (\gamma t ) 
+\frac{1}{2}\left[ {\bar n}_{\rm R} (1+2{\bar n}_{\rm R} )-\frac{1}{2} |\alpha|^2 \right] (\gamma t )^2 
\notag   \\ 
& +{\rm O}\left[ (\gamma t )^3 \right],      \qquad         (\gamma t \ll 1),
\label{F1st}
\end{align} 
as well as its asymptotic behavior:
\begin{align}
& {\cal F}^{(1)}(t)=\frac{1}{ 1+{\bar n}_{\rm R} } 
\exp\left( -\frac{ |\alpha|^2 }{ 1+{\bar n}_{\rm R} } \right)      
\left[ 1+2\, \frac{ |\alpha|^2 }{ 1+{\bar n}_{\rm R} }\sqrt{\eta}   \right.  \notag   \\
& \left. +{\rm O}(\eta) \right],         \qquad         (\gamma t \gg 1).
\label{F1lt}
\end{align}  
From Eq. (\ref{F1st}) one gets the rate of dechorence of the field mode:
$\, {\Gamma}_{\rm d}^{(1)}:=-\dot{\cal F}^{(1)}(0)={\bar n}_{\rm R} \gamma $; its inverse
is the decoherence time $ {\tau}_{\rm d}^{(1)}=1/( {\bar n}_{\rm R} \gamma).$

\subsubsection{Purity}

The purity (\ref{Pcoh}) of the transient DTS (\ref{DTS}) does not depend on the coherent 
amplitude $\alpha$. It is a strictly monotonic and strictly convex function of time, 
which decreases from the initial pure-state value ${\cal P}(0)=1$ towards the steady-state limit 
${\cal P}(\infty)=( 1+2{\bar n}_{\rm R} )^{-1.}$ 

We write down an expansion of the purity (\ref{Pcoh}) that is valid at short times: 
\begin{align}
& {\cal P}(t)=1-2{\bar n}_{\rm R} (\gamma t ) 
+\frac{1}{2}\left[ 2{\bar n}_{\rm R} (1+4{\bar n}_{\rm R} ) \right] (\gamma t )^2 
+{\rm O}\left[ (\gamma t )^3 \right],      \notag   \\         
& (\gamma t \ll 1),
\label{Pst}
\end{align} 
and the asymptotic formula:
\begin{align}
& {\cal P}(t)=\frac{1}{ 1+2{\bar n}_{\rm R} } \left[ 1+ \frac{ 2{\bar n}_{\rm R} }
{ 1+2{\bar n}_{\rm R} }\, \eta +{\rm O}\left( {\eta}^2 \right) \right],        \notag   \\
& (\gamma t \gg 1).
\label{Plt}
\end{align}  
According to Eq. (\ref{Pst}), the rate of mixing of the initially coherent state is
$\, {\Gamma}_{\rm m}:=-\dot{\cal P}(0)=2{\bar n}_{\rm R} \gamma $; its inverse defines 
a conventional mixing time: $ {\tau}_{\rm m}=1/( 2{\bar n}_{\rm R} \gamma).$

On the one hand, from Eqs. (\ref{F1st}) and (\ref{Pst}), we get the short-time expression
of the HS evolution distance (\ref{GC1}): 
\begin{equation}
{\cal G}^{(1)}(t)=\dot{\cal G}^{(1)}(0) \, t +{\rm O}\left[ (\gamma t )^2 \right], 
\qquad          ( \gamma t \ll 1),
\label{G1st}
\end{equation}
with the slope at the origin
\begin{equation}
\dot{\cal G}^{(1)}(0)=\sqrt{2} \gamma \left( {\bar n}_{\rm R}^2 
+ \frac{1}{4} |\alpha|^2 \right)^{\frac{1}{2} }.
\label{dotG1(0)}
\end{equation}
On the other hand, Eqs. (\ref{F1lt}) and (\ref{Plt}) provide the steady-state sign rule:
\begin{align} 
& \text{sgn}\left\{ \lim_{t \to \infty} \left[ {\cal F}^{(1)}(t)-{\cal P}(t) \right] \right\}           \notag  \\
& =\text{sgn}  \left[ ( 1+{\bar n}_{\rm R} ) \ln \left( 1+\frac{ {\bar n}_{\rm R} }
{ 1+{\bar n}_{\rm R} } \right) -|\alpha|^2 \right].      
\notag      
\end{align}

\subsubsection{Upper bound for the speed of evolution}

In the sequel, we employ two approximations of the exact formula (\ref{vcoh^2})
obtained by keeping only its most important term:
\begin{align}
& {\tilde v}(t) = \sqrt{2} \left[ {\omega}^2 +\left( \frac{1}{2} \gamma \right)^2 \right]^{\frac{1}{2} } 
\frac{ |\alpha| \, \sqrt{\eta} }{ 1+2{\bar n}_{\rm T}(t)} \, ,      \notag  \\ 
& ( |\alpha| \omega \gg {\bar n}_{\rm R} \gamma);
\label{tv1}
\end{align}
\begin{equation}
{\tilde v}(t) = \sqrt{2} \frac { {\bar n}_{\rm R} \gamma \eta } 
{\left[ 1+2{\bar n}_{\rm T}(t) \right]^{\frac{3}{2} } } ,
\qquad        ( |\alpha| \omega \ll {\bar n}_{\rm R} \gamma).
\label{tv2}
\end{equation}
The condition for the validity of Eq. (\ref{tv1}) is largely fulfilled. For instance,
when choosing $|\alpha|=10^{-3}$ and at room temperature, it holds for laser frequencies
ranging from UV to IR. Therefore, Eq. (\ref{tv2}) may only be used for a low-intensity laser
whose frequency is in the far IR. Note that Eq. (\ref{tv1}) becomes exact for a dissipating
coherent field mode $( \alpha \neq 0, \,{\bar n}_{\rm R}=0),\,$ while Eq. (\ref{tv2}) 
is rigorously exact when describing the thermalization of the vacuum 
$(\alpha =0, \, {\bar n}_{\rm R} >0).$

\subsection{Quantum speed limit times}

\subsubsection{Exact general formulas}

When using the purity (\ref{Pcoh}) of the transient DTS, we find the following explicit 
expression of the QSL (\ref{QSL1}): 
\begin{align}
& \overline{v_{\cal F} } (t)=v_{\cal F}(0) \overline{\sqrt{ {\cal P} } }(t) 
=\frac{ v_{\cal F}(0) }{\left( 1+2{\bar n}_{\rm R} \right)^{\frac{1}{2} } }     \notag   \\
& \times \left\{ 1+\ln\left( 1+\frac{-1+[1+2{\bar n}_{\rm T}(t) ]^{\frac{1}{2} } }
{1+\left( 1+2{\bar n}_{\rm R} \right)^{\frac{1}{2} } }\right) \right\}.
\label{v(t)coh}
\end{align}
In the above formula, the QSL $v_{\cal F}(0)$, Eq. (\ref{v(0)coh}), is the only factor that depends 
on $|\alpha|^2$. Let us write down the short-time approximation of the QSL (\ref{v(t)coh}):
\begin{align}
& \overline{v_{\cal F} } (t)=v_{\cal F}(0) \left\{ 1-\frac{1}{2} {\bar n}_{\rm R} (\gamma t)
+\frac{1}{6} {\bar n}_{\rm R} \left( 1+3{\bar n}_{\rm R} \right) (\gamma t)^2      \right.   \notag   \\
& \left.  +{\rm O}\left[ (\gamma t)^3 \right] \right\},      \qquad          (\gamma t \ll 1),
\label{v(t)st}
\end{align}
as well as its long-time behavior:
\begin{align}
& \overline{v_{\cal F} } (t)= \frac{ v_{\cal F}(0) }{\left( 1+2{\bar n}_{\rm R} \right)^{\frac{1}{2} } }   
\left\{ 1+\frac{2}{\gamma t } \ln\left[ \frac{2}{1+(1+2{\bar n}_{\rm R})^{-\frac{1}{2} } } \right]
\right.   \notag   \\
& \left.  -\frac{ {\bar n}_{\rm R} }{1+2{\bar n}_{\rm R} } \frac{1}{\gamma t } 
\, {\rm O}(\eta) \right\},         \qquad          (\gamma t \gg 1).
\label{v(t)lt}
\end{align}

\subsubsection{Approximate formulas for $\alpha \neq 0$ }

Further, we employ the approximate formulas (\ref{tv1}) and (\ref{tv2}) 
to obtain two alternative QSLs (\ref{QSL2}): 
\begin{align}
& \overline{\tilde v} (t) = \sqrt{2} \left[ \frac{1+\left( \frac{\gamma }{2\omega } \right)^2 } 
{ (1+2{\bar n}_{\rm R} ) 2{\bar n}_{\rm R} } \right]^{\frac{1}{2} } 
\frac{\omega |\alpha|}{ \gamma t }       \notag   \\
& \times \left\{ 2\ln\left[ (1+2{\bar n}_{\rm R} )^{\frac{1}{2} }+( 2{\bar n}_{\rm R} )^{\frac{1}{2} }\right] 
\right.   \notag   \\
& \left. +2\ln\left[ (1+2{\bar n}_{\rm R} )^{\frac{1}{2} }-( 2{\bar n}_{\rm R} \eta)^{\frac{1}{2} }\right] 
\right.   \notag   \\
& \left. -\ln\left[ 1+2{\bar n}_{\rm T}(t) \right] \right\},
\qquad      ( |\alpha| \,\omega \gg {\bar n}_{\rm R} \gamma),
\label{btv1}
\end{align}
and, respectively,
\begin{align}
& \overline{ {\tilde v} }(t)=\frac{\sqrt{2} }{t} 
\left\{ 1- \left[ 1+2{\bar n}_{\rm T}(t) \right]^{-\frac{1}{2} } \right\},         \notag   \\   
& ( |\alpha| \, \omega \ll {\bar n}_{\rm R} \gamma).
\label{btv2}
\end{align}

The QSL (\ref{btv1}) has the following short-time behavior: 
\begin{align}
& \overline{\tilde v} (t) = \sqrt{2} \left[ 1+\left( \frac{\gamma }{2\omega } \right)^2  
\right]^{\frac{1}{2} } \omega |\alpha|            \notag   \\
& \times  \left\{ 1- \frac{1}{4} \left( 1+4{\bar n}_{\rm R} \right) (\gamma t )
+ \frac{1}{24} \left( 1+16\, {\bar n}_{\rm R} +32\, {\bar n}_{\rm R}^2 \right) (\gamma t )^2    
\right.      \notag   \\
& \left. +{\rm O}\left[ (\gamma t)^3 \right] \right\},        \qquad                  
( |\alpha| \, \omega \gg {\bar n}_{\rm R} \gamma , \;\;\; \gamma t \ll 1),
\label{btv1st}
\end{align}
while it vanishes in the steady-state regime:
\begin{align}
& \overline{\tilde v} (t) = \sqrt{2} \left[ \frac{1+\left( \frac{\gamma }{2\omega } \right)^2 } 
{ (1+2{\bar n}_{\rm R} ) 2{\bar n}_{\rm R} } \right]^{\frac{1}{2} } 
\frac{\omega |\alpha|}{ \gamma t }       \notag   \\
& \times \left\{ 2\ln\left[ (1+2{\bar n}_{\rm R} )^{\frac{1}{2} }+( 2{\bar n}_{\rm R} )^{\frac{1}{2} }\right]  -2\left( \frac{ 2{\bar n}_{\rm R} }{1+2{\bar n}_{\rm R} }\, \eta \right)^{\frac{1}{2} } 
\right.      \notag   \\
& \left. +{\rm O}\left( {\eta}^{\frac{3}{2} } \right)  \right\},      
\quad     ( |\alpha| \omega \gg {\bar n}_{\rm R} \gamma , \;\;\; \gamma t \gg 1 ).
\label{btv1lt}
\end{align}
For a dissipating coherent field, $({\bar n}_{\rm R}=0)$, the formula (\ref{btv1})
becomes exact, simplifying to:
\begin{align}
& \overline{\tilde v} (t) = \sqrt{2}\, \left[ 1+\left( \frac{\gamma }{2\omega } \right)^2  
\right]^{\frac{1}{2} } \frac{\omega |\alpha|}{ \gamma t } \, 2\left( 1-\sqrt{\eta} \right),       \notag   \\
& ( {\bar n}_{\rm R}=0).
\label{n_R=0}
\end{align}

In turn, the short-time behavior of the QSL (\ref{btv2}) reads:
\begin{align}
& \overline{\tilde v} (t) = \sqrt{2}\, {\bar n}_{\rm R} \gamma 
\left\{ 1- \frac{1}{2} \left( 1+3{\bar n}_{\rm R} \right) (\gamma t )     \right.      \notag    \\  
& \left. + \frac{1}{6} \left( 1+9{\bar n}_{\rm R} +15\, {\bar n}_{\rm R}^2 \right) (\gamma t )^2    
+{\rm O}\left[ (\gamma t)^3 \right] \right\},        \notag    \\            
& ( |\alpha| \, \omega \ll {\bar n}_{\rm R} \gamma , \;\;\; \gamma t \ll 1),
\label{btv2st}
\end{align}
As expected, it has a vanishing asymptotic limit:
\begin{align}
& \overline{\tilde v} (t) =\frac{\sqrt{2} }{t} \left[ 1-(1+2{\bar n}_{\rm R} )^{-\frac{1}{2} }   
\left( 1+ \frac{ {\bar n}_{\rm R} }{1+2{\bar n}_{\rm R} } \eta \right)    \right.     \notag   \\
& \left. +{\rm O}\left( {\eta}^2 \right) \right],      \qquad        
( |\alpha| \, \omega \ll {\bar n}_{\rm R} \gamma, \;\;\; \gamma t \gg 1 ).
\label{btv2lt}
\end{align}

The new QSLTs are similar to the old ones, Eqs. (\ref{tauF}),  (\ref{ttauF}), and (\ref{ttauG}), 
\begin{align} 
{\tau}_{\cal F}^{(1)}(t):&= \frac{1-{\cal F}^{(1)}(t) }{\overline{v_{\cal F} }(t) }     \notag     \\
& \geqq  {\tau}_{\cal F}^{ (1){\rm min} }(t) = \frac{1-{\cal F}^{(1)}(t) }{v_{\cal F}(0) },
\label{tau1F}
\end{align}
\begin{equation} 
{\tilde \tau}_{\cal F}^{(1)}(t):= \frac{1-{\cal F}^{(1)} (t) }{\overline{ \tilde{v} }(t) },
\label{ttau1F}
\end{equation}
\begin{equation} 
{\tilde \tau}_{\cal G}^{(1)}(t):= \frac{ {\cal G}^{(1)} (t) }{\overline{ \tilde{v} }(t) },
\label{ttau1G}
\end{equation}
but are comparatively smaller. Owing to the inequality ${\cal F}^{(1)}(t) \leqq  \sqrt{{\cal P}(t)}$
and to its consequence,   $${\cal G}^{(1)}(t)\geqq 1-{\cal F}^{(1)}(t),$$ 
we get a relation analogous to the general formula (\ref{ttauGF}): 
\begin{equation}
\tilde{\tau}_{\cal G}^{(1)}(t)  \geqq  \tilde{\tau}_{\cal F}^{(1)}(t).
\label{ttau1GF}
\end{equation}
In Fig.1, the above QSLTs,  Eqs. (\ref{tau1F})-(\ref{ttau1G}), are visualized for the dissipation 
$(\bar n_{\rm R}=0)$ of an input coherent field with amplitude $\alpha=2$. 
For convenience, we choose a ratio $\gamma/\omega=0.1.$

\begin{figure}[h]
\center
\includegraphics[width=9cm]{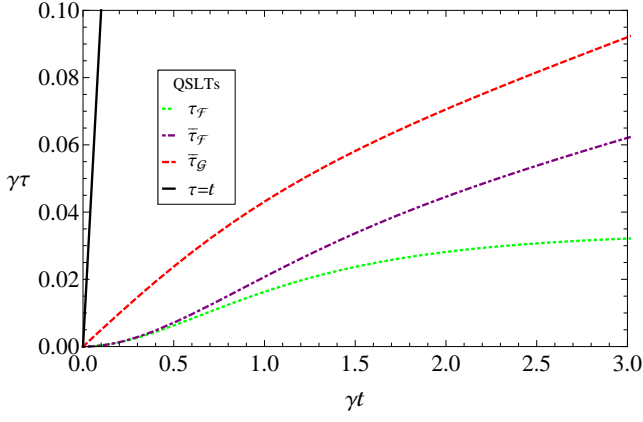}
\caption{(Color online)   The QSLTs  ${\tau}_{\cal F}^{ (1){\rm min} }(t)$, Eq.(5.29) 
(dotted green line), ${\tilde \tau}_{\cal F}^{(1)}(t)$, Eq.(5.30) (dotted-dashed purple line), 
and ${\tilde \tau}_{\cal G}^{(1)}(t)$, Eq.(5.31) (dashed red line) for a coherent field 
of amplitude $\alpha=2$, evolving in contact with a zero-temperature reservoir 
under the condition  $\gamma/\omega=0.1.$ Comparison with the actual time of evolution 
(solid black line) shows that these lower bounds are not tight.}
\end{figure}

\subsubsection{Exact formulas in the limiting case $\alpha =0$ }

When one deals with thermalization of the vacuum 
$(\alpha =0, \, {\bar n}_{\rm R} >0)$,
all five functions (\ref{F1,-1})-(\ref{FT}) coincide with the exact fidelity 
of evolution (\ref{Fcoh}), which is denoted ${\cal F}_0(t)$ and reads:
\begin{equation}
{\cal F}_0(t)=\frac{1}{ 1+{\bar n}_{\rm T}(t) }.
\label{F_0}
\end{equation}
On the other hand, the QSL (\ref{v(0)coh}) simplifies to:
\begin{equation}
v_{\cal F}(0;0) =\gamma \left[ 1+2{\bar n}_{\rm R} (1+{\bar n}_{\rm R} ) \right]^{\frac{1}{2} }.            
\label{vF(0)}
\end{equation}
Further, the purity ${\cal P}_0(t)$ has the general expression (\ref{Pcoh}), 
while the approximate QSL (\ref{btv2}) becomes exact: 
\begin{align}
\overline{ {\tilde v} }_0(t)=\frac{ \sqrt{2} }{t} \,
\frac{ \left[ 1+2{\bar n}_{\rm T}(t) \right]^{\frac{1}{2} } -1 }
{ \left[ 1+2{\bar n}_{\rm T}(t) \right]^{\frac{1}{2} } }.
\label{btv0}
\end{align}
The HS measure of evolution (\ref{HS1}),
\begin{equation} 
{\cal G}_0(t)=\left[ 1+ {\cal P}_0(t)-2 {\cal F}_0(t) \right]^{\frac{1}{2} },
\label{G_0}
\end{equation}
has the expression:
\begin{equation} 
{\cal G}_0(t)=\frac{ \sqrt{2} \, {\bar n}_{\rm T}(t) }{\left[ 1+2{\bar n}_{\rm T}(t) \right]^{\frac{1}{2} }
\left[ 1+{\bar n}_{\rm T}(t) \right]^{\frac{1}{2} } }.
\label{G0}
\end{equation}
Hence, we get the fidelity-based QSLTs:
\begin{align}
& {\tau}_{\cal F}^{\rm min}(0;t)=\frac{1-{\cal F}_0(t) }{v_{\cal F}(0;0) }          \notag  \\
& =\frac{1}{\gamma} \frac{ {\bar n}_{\rm T}(t) }{ 1+{\bar n}_{\rm T}(t) } 
\left[ 1+2{\bar n}_{\rm R} (1+{\bar n}_{\rm R} ) \right]^{-\frac{1}{2} },  
\label{tauF0}
\end{align}
\begin{align}
& \tilde{\tau}_{\cal F}(0; t)=\frac{1-{\cal F}_0(t) }{\overline{ {\tilde v} }_0(t) }   \notag  \\
& =\left[ \frac{ 1+2{\bar n}_{\rm T}(t) }{ 2+2{\bar n}_{\rm T}(t) } \right]^{\frac{1}{2} }    
\frac{1+\left[ 1+2{\bar n}_{\rm T}(t) \right]^{\frac{1}{2} } }
{ 2\left[ 1+{\bar n}_{\rm T}(t) \right]^{\frac{1}{2} } }\, t,
\label{ttauF0}
\end{align}
as well as the HS-metric-based one,
\begin{equation}
\tilde{\tau}_{\cal G}(0; t)=\frac{ {\cal G}_0(t) }{\overline{ {\tilde v} }_0(t) }
=\frac{ 1+\left[ 1+2{\bar n}_{\rm T}(t) \right]^{\frac{1}{2} } } 
{ 2\left[ 1+{\bar n}_{\rm T}(t) \right]^{\frac{1}{2} } }\, t.
\label{ttauG0}
\end{equation}

\begin{figure}[h]
\center
\includegraphics[width=9cm]{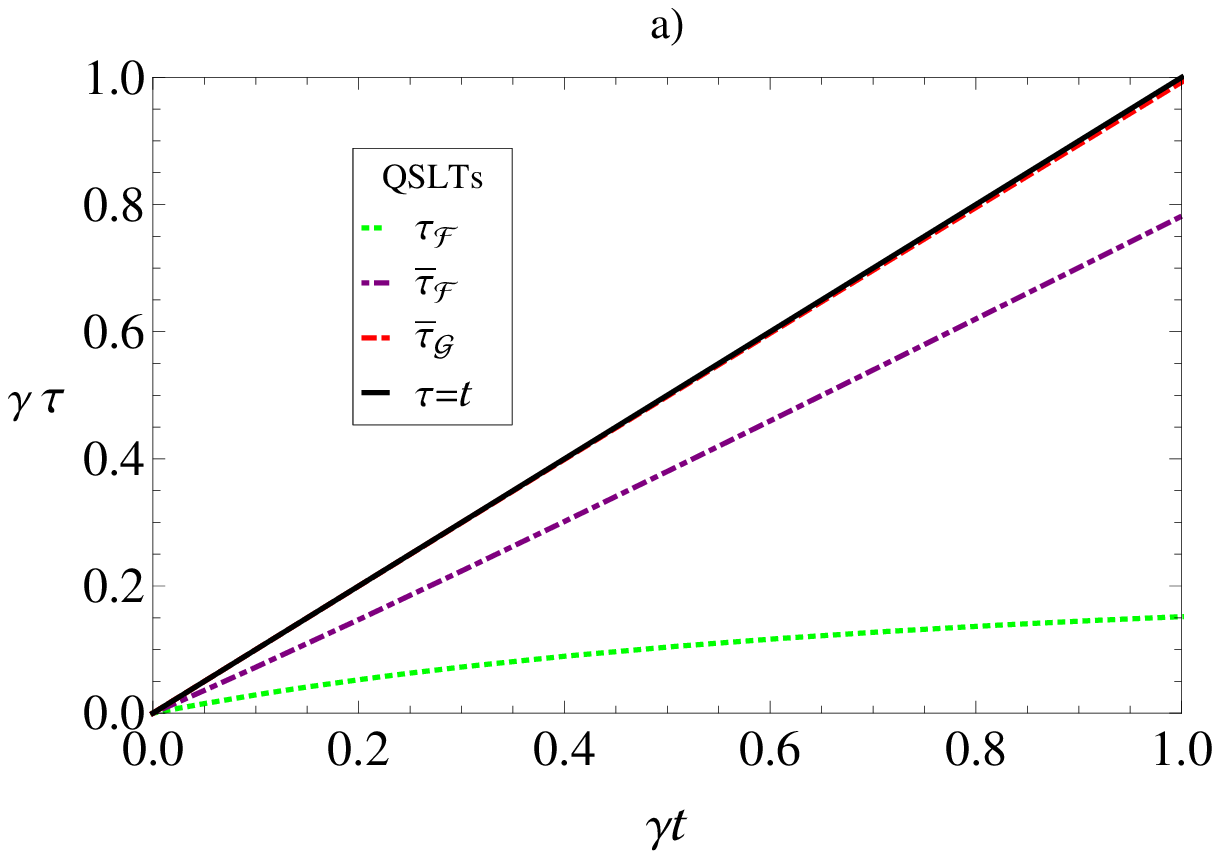}
\includegraphics[width=9cm]{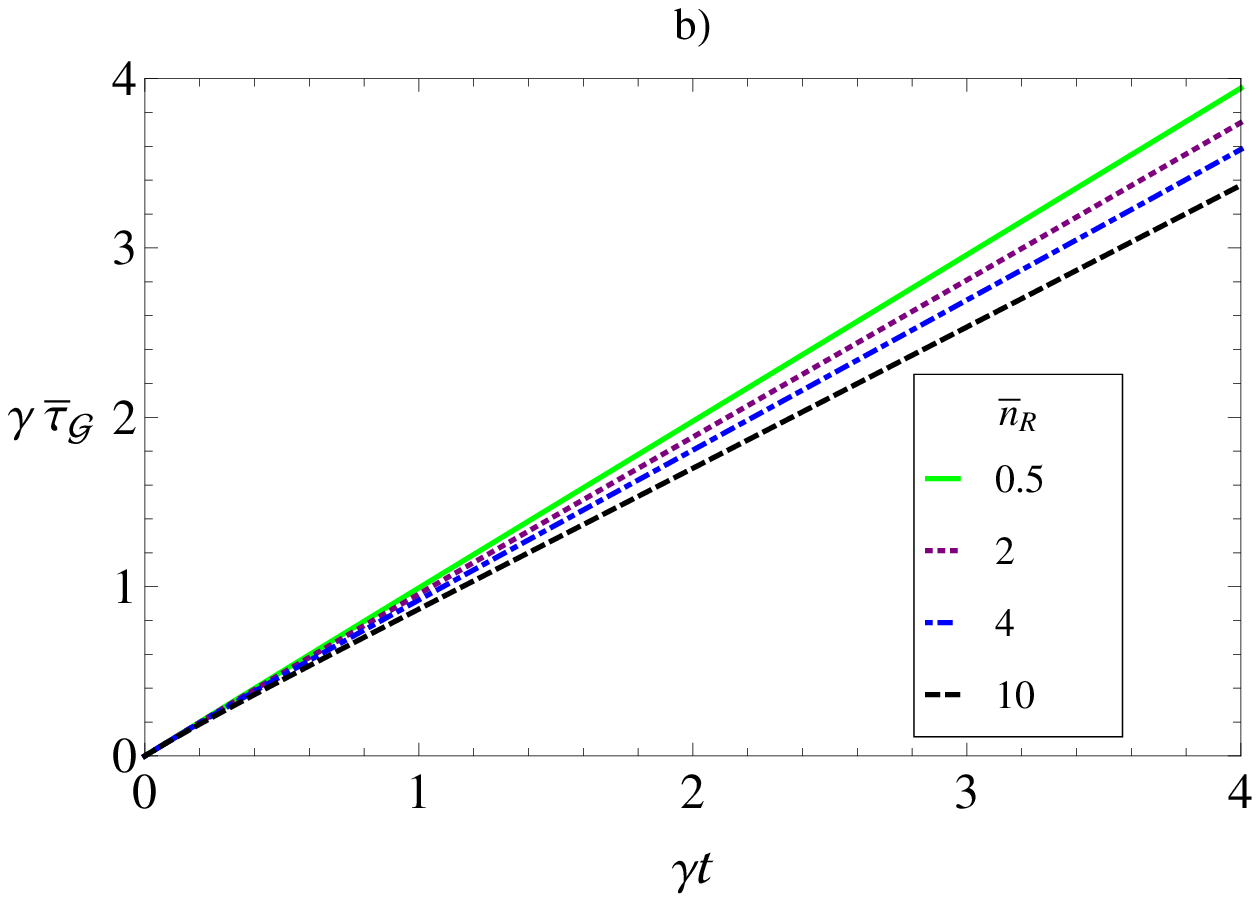}
\caption{(Color online)  a) The QSLTs (\ref{tauF0}) (dotted green line), 
(\ref{ttauF0}) (dotted-dashed purple line), and (\ref{ttauG0}) (dashed red line)
for thermalization of the vacuum $(\alpha=0, \, {\bar n}_{\rm R}=0.5)$ compared 
with the actual time of evolution (solid black line).
b) The Hilbert-Schmidt QSLT (\ref{ttauG0}) remains tight even for noisier thermal 
environments. From top to the bottom:  ${\bar n}_{\rm R}=0.5$ (solid green line),
${\bar n}_{\rm R}=2$ (dotted purple line),  ${\bar n}_{\rm R}=4$ (dotted-dashed blue line),
and ${\bar n}_{\rm R}=10$ (dashed black line).}
\end{figure}

Figure 2 a) displays the expected hierarchy of the QSLTs (\ref{tauF0})-(\ref{ttauG0}).
The HS QSLT (\ref{ttauG0}) is so close to the actual driving time $t$ that their graphs 
are practically superposed when $\gamma t \leqq 1$. Its tightness is quite remarkable 
even for higher values of the Bose-Einstein mean photon occupancy ${\bar n}_{\rm R}$,
as illustrated in Fig. 2 b).

Since the field mode is fed only by the reservoir, all photons are thermal:
$\langle {\hat a}^{\dag}\hat a \rangle(t) = {\bar n}_{\rm T}(t)$, Eq. (\ref{nT}). 
In the twofold limiting case ${\bar n}_{\rm R}=0$, there are no photons in the cavity, 
while all atoms could be trapped to form a Bose-Einstein condensate in its ground-state 
configuration. The absence of any interaction means that the field-reservoir 
coupling constant vanishes: $\gamma=0$. Consequently, there is no field-state evolution 
and both QSLs (\ref{vF(0)}) and (\ref{btv0}) vanish: 
$$v_{\cal F}(0;0)=0,   \qquad   \overline{ {\tilde v} }_0(t)=0,  \qquad   ({\bar n}_{\rm R}=0).$$ 

\section{Thermalization of a Fock state}

\subsection{Transient state}

The CF (\ref{CF}) of an initial number state $| M\rangle \langle M |$ is
\begin{align} 
& {\chi}_M (\lambda, {\lambda}^{\ast}, 0))
:=\langle M | {\hat D} (\lambda, {\lambda}^{\ast} )| M \rangle        \notag    \\ 
& = \exp\left( -\frac{1}{2} |\lambda|^2 \right) L_M \left( |\lambda|^2 \right),
\label{CFM0}
\end{align}
where $L_M(x)$ stands for the Laguerre polynomial of degree $M$, Eq. (\ref{1F1}).
According to Eqs. (\ref{NCFt}), (\ref{lambda}), and (\ref{CF}),  the CF and the NCF 
of the thermalized number state ${\hat \rho}_M(t)$ are given by the formula::
\begin{align}
& \chi_M(\lambda, {\lambda}^{\ast}, t)=\exp\left( -\frac{1}{2} |\lambda|^2 \right)
\chi_M^{(N)}(\lambda, {\lambda}^{\ast}, t)                          \notag    \\
& =\exp\left( -\left[ \frac{1}{2}+\bar{n}_{\rm T}(t) \right] |\lambda|^2 \right)         
L_M \left( \eta \, |\lambda|^2 \right).
\label{CFMt}
\end{align}
Therefore, the transient state ${\hat \rho}_M(t)$ describes the linear superposition
of an attenuated mode with a definite initial number of photons $M$ 
and a thermal mode whose mean photon number $\bar{n}_{\rm T}(t)$,  Eq. (\ref{nT}),
is fed by the reservoir.. 

The Fourier transforms of the CF and the NCF in Eq. (\ref{CFMt}) are respectively 
the Wigner function $W_M(\beta, {\beta}^{\ast}, t)$ and, up to a factor $\pi,$
the Glauber-Sudarshan $P$ function  $\pi P_M(\beta, {\beta}^{\ast}, t)$ \cite{CG}.
In Ref. \cite{PIT13}, these functions are evaluated and found to have a similar structure:
\begin{align}
& W_M(\beta, {\beta}^{\ast}, t)=\frac{2}{1+2{\bar n}_{\rm T}(t) }        
\exp\left[ -\frac{2|\beta|^2}{1+2{\bar n}_{\rm T}(t) } \right]              \notag    \\ 
& \times \left[ \frac{1+2{\bar n}_{\rm R} -2(1+{\bar n}_{\rm R} ) \eta}
{1+2{\bar n}_{\rm T}(t) }\right]^M          \notag    \\ 
& \times L_M\left( -\frac{ 4\eta \, |\beta|^2  }{\left[ 1+2{\bar n}_{\rm T}(t) \right] 
\left[ 1+2{\bar n}_{\rm R} -2(1+{\bar n}_{\rm R} ) \eta \,\right] } \right) ;
\label{WM}
\end{align}
\begin{align}
& P_M(\beta, {\beta}^{\ast}, t)=\frac{1}{\pi} \frac{1}{ {\bar n}_{\rm T}(t) }        
\exp\left[ -\frac{ |\beta|^2 }{ {\bar n}_{\rm T}(t) } \right]              \notag    \\ 
& \times \left[ \frac{ {\bar n}_{\rm R} -(1+{\bar n}_{\rm R} ) \eta}
{ {\bar n}_{\rm T}(t) } \right]^M          \notag    \\ 
& \times L_M\left( -\frac{ \eta \, |\beta|^2) }{ {\bar n}_{\rm T}(t) 
\left[ {\bar n}_{\rm R} -(1+{\bar n}_{\rm R} ) \eta \,\right] } \right) .
\label{PM}
\end{align}

In what follows, we leave out the special case $M=0$, \\ since thermalization of the vacuum 
is already treated in the preceding section. For $M>0,$ the regular functions  (\ref{WM}) 
and (\ref{PM}) are positive on the whole phase space if the arguments of their Laguerre 
polynomials are negative:
\begin{equation}
t \geqq t_{\rm w}:=\frac{1}{\gamma} \ln\left( 1+\frac{1}{1+2{\bar n}_{\rm R} }\right) 
\Longrightarrow W_M(\beta, {\beta}^{\ast}, t) >0;    
\label{WM>0}
\end{equation}
\begin{equation}
t \geqq t_{\rm c}:=\frac{1}{\gamma} \ln\left( 1+\frac{1}{ {\bar n}_{\rm R} }\right) 
\Longrightarrow P_M(\beta, {\beta}^{\ast}, t) >0.    
\label{PM>0}
\end{equation}
As shown in Ref. \cite{PIT13}, the damping of an excited number state involves 
three successive stages, each of them with specific features of the evolved state:
\begin{itemize}
\item{ $\; t \in [\, 0,  t_{\rm w} ).$ {\em Strong non-classicality:} Both functions 
$W_M(\beta, {\beta}^{\ast}, t)$ and $P_M(\beta, {\beta}^{\ast}, t)$ have some negative values; }
\item{ $\; t \in [\, t_{\rm w}, t_{\rm c} ).$ {\em Weak non-classicality:}  
$W_M(\beta, {\beta}^{\ast}, t) >0$, while $P_M(\beta, {\beta}^{\ast}, t)$ 
is negative somewhere; }
\item{ $\; t \geqq  t_{\rm c}.$ {\em Classicality:} $(2\pi)^{-1}W_M(\beta, {\beta}^{\ast}, t) >0 \;\;
\text{and} \\ 2^{-1}P_M(\beta, {\beta}^{\ast}, t) >0,$ } so that both functions are genuine 
densities of probability in the phase space.
\end{itemize}
The threshold times $t_{\rm w}$ and $t_{\rm c}$  do not depend on the initial number 
of photons $M$. We term them the weak non-classicality threshold and the classicality one, 
respectively. Remark that the transient state ${\hat \rho}_M(t)$ of a dissipating mode 
$( {\bar n}_{\rm R}=0 )$ is always non-classical because $t_{\rm c}=\infty ;$ on the contrary, 
its weak non-classicality threshold is finite: $t_{\rm w}=\frac{1}{\gamma} \ln(2).$

\subsection{Main ingredients}

For initial Fock states, we write the specific versions of the general formulas
(\ref{fidev1})-(\ref{v^2}), where the initial CF (\ref{CFM0}) and, respectively
the transient one (\ref{CFMt}) should be substituted.

\subsubsection{Fidelity of evolution}

The fidelity of evolution (\ref{fidev1}) is evaluated making use of the integral (\ref{LTLL2}):
\begin{align}
{\cal F}_M(t)= \frac{ {\eta}^M \, [u(t)]^{-M} } { [ 1+{\bar n}_{\rm T}(t) ]^{2M+1} }                 
\, {_{2}F_{1} }\left(-M, -M; \,1; \, u(t) \right).
\label{F_M}
\end{align}
In Eq. (\ref{F_M}), the Gauss hypergeometric function $ {_{2}F_{1} }$, Eq. (\ref{2F1}),
is a monic polynomial of degree $M$ with the positive variable
\begin{equation}
u(t):=\frac{1}{ {\bar n}_{\rm R} ( 1+{\bar n}_{\rm R} ) } \, \frac{\eta}{ (1-\eta )^2 },          
\qquad        ( {\bar n}_{\rm R} >0).
\label{u(t)}
\end{equation}
This variable is a strictly decreasing and strictly convex function of time: it decreases 
from $u(0)=\infty$ to $u(\infty)=0$. The above-mentioned property of the hypergeometric 
polynomial $ {_{2}F_{1} }$ in Eq. (\ref{F_M}) enables us to write the formula:
\begin{equation}
\lim_{u \to \infty} \left[ u^{-M} {_{2}F_{1}}\left(-M, -M; \,1; \, u \right) \right] =1.
\end{equation}
Therefore, in the limit case ${\bar n}_{\rm R}=0$, corresponding to the dissipation 
of the mode, Eq. (\ref{F_M}) simplifies to:
\begin{equation}
{\cal F}_M(t)={\eta}^M,     \qquad        ( {\bar n}_{\rm R} =0).
\label{F_Mn=0}
\end{equation}

The time derivative of the fidelity (\ref{F_M}) has a rather complicated expression:
\begin{align}
& \dot{\cal F}_M(t)= -\gamma {\cal F}_M(t) \left\{ M+(2M+1) \frac{ {\bar n}_R \, \eta}
{1+{\bar n}_T(t) }     \right.         \notag \\ 
& \left.+ \left[ -M+M^2 u(t) \,\frac{ {_{2}F_{1}}\left(-M+1, -M+1; \,2; \,u(t) \right) }
{ {_{2}F_{1}}\left(-M, -M; \,1; \, u(t) \right) } \right]    \right.         \notag \\ 
& \left. \times \frac{1+\eta}{1-\eta} \right\}.
\label{dotF_M}
\end{align}
For that very reason, it is hard to answer the general question whether the function of time 
(\ref{F_M}) is monotonic or not, whatever values of its parameters $M$ and ${\bar n}_R$ 
are chosen. Nevertheless, in the simplest case $M=1$, a thorough analysis of damping  
is carried out in Appendix C.

The fidelity of evolution (\ref{F_M}) has the short-time expansion:
\begin{align}
& {\cal F}_M(t)=1-\left[ M+(2M+1) {\bar n}_{\rm R} \right] (\gamma t)                         \notag   \\  
& +\frac{1}{2} \left[ M^2 +\left( 6M^2+4M+1 \right) {\bar n}_{\rm R}         \right.          \notag   \\
& \left. +2\left( 3M^2+3M+1 \right) {\bar n}_{\rm R}^2 \right] (\gamma t)^2                 \notag   \\
& +{\rm O}\left[ (\gamma t)^3 \right],      \qquad          (\gamma t \ll 1),
\label{F_Mst}
\end{align}
Accordingly, the modulus of the slope of its graph at $t=0$, i. e.,  the rate of decoherence
of the initial Fock state is ${\Gamma}_{\rm d}(M, {\bar n}_{\rm R} ):
=-\dot{\cal F}_M(0)=\gamma \left[ M+(2M+1) {\bar n}_{\rm R} \right] $.

We also write the asymptotic formula:
\begin{align}
& {\cal F}_M(t)= \frac{1}{1+{\bar n}_R }\left( \frac{ {\bar n}_R }{1+{\bar n}_R } \right)^M
\left[ 1+ \frac{ (M-{\bar n}_R )^2 }{ {\bar n}_R (1+{\bar n}_R ) }\, \eta  \right.  \notag  \\
& \left. + \frac{1}{2\left[ {\bar n}_R (1+{\bar n}_R ) \right]^2 } \left( \frac{1}{2} [M(M-1)]^2
+2{\bar n}_R \left\{ M{\bar n}_R             \right. \right. \right.         \notag  \\
& \left. \left. \left.  + (M- {\bar n}_R )\left[ -2M(M-1)+{\bar n}_R (3M- {\bar n}_R ) \right]
\right\} \right) {\eta}^2       \right.      \notag  \\
& \left. +{\rm O}\left( {\eta}^3 \right) \right],   \qquad   ( {\bar n}_R >0, \;\;  \gamma t \gg 1).
\label{F_Mlt}
\end{align}
From Eq. (\ref{F_Mlt}) one learns that the steady-state limit
\begin{equation}
\lim_{t \to \infty}[{\cal  F}_M(t)]= \frac{1}{1+{\bar n}_R }
\left( \frac{ {\bar n}_R }{1+{\bar n}_R } \right)^M
\label{FMas}
\end{equation}
is always reached from above. In fact, this limit is an input-output fidelity, 
i. e., the fidelity between the initial Fock state 
$| M\rangle \langle M |$ and the TS ${\hat \rho}_{\rm T}({\bar n}_{\rm R} )$, 
which is eventually imposed by the bosonic reservoir to the field mode at equilibrium: 
$$\lim_{t \to \infty}[{\cal  F}_M(t)]=\langle M |{\hat \rho}_{\rm T}({\bar n}_{\rm R} )| M \rangle.$$

At the classicality threshold $t_{\rm c}$, Eq. (\ref{PM>0}), the variable (\ref{u(t)}) 
is equal to one, 
$$ u(t)=1 \; \iff  \; t= t_{\rm c},      \qquad        ( {\bar n}_{\rm R} >0). $$
Consequently, the fidelity (\ref{F_M}), as well as its time derivative (\ref{dotF_M}) reduce
to a single monomial owing to Gauss's summation formula (\ref{Gauss}): 
\begin{equation}
{\cal F}_M(t_{\rm c})=\binom{2M}{M} \, \frac{ ({\bar n}_{\rm R})^M
(1+{\bar n}_{\rm R})^{M+1} }{ \left( 1+2{\bar n}_{\rm R} \right)^{2M+1} },    
\;\; ( {\bar n}_{\rm R} >0);
\label{FMG}
\end{equation}
\begin{equation}
{\dot{\cal F} }_M(t_{\rm c} )=-\gamma \frac{ M+2{\bar n}_{\rm R}^2 }{ 2(1+2{\bar n}_{\rm R} )} 
{\cal F}_M(t_{\rm c} )<0,   \;\;   ( {\bar n}_{\rm R} >0).
\label{dotFMG}
\end{equation}
Both fidelities (\ref{FMas}) and (\ref{FMG}) decrease with the initial number of photons $M$.

\subsubsection{Purity}

We perform the integral (\ref{pur})  by employing once more Eq. (\ref{LTLL2}) 
to get the evolved purity: 
\begin{align}
{\cal P}_M(t)= \frac{ {\eta}^{2M} \, [w(t)]^{-M} } { [ 1+2{\bar n}_{\rm T}(t) ]^{2M+1} }                 
\, {_{2}F_{1} }\left(-M, -M; \,1; \, w(t) \right).
\label{P_M}
\end{align}
Here we have introduced the positive variable
\begin{equation}
w(t):=\left( \frac{1}{1+2{\bar n}_{\rm R} }\, \frac{\eta}{1-\eta} \right)^2,        
\label{w(t)}
\end{equation}
which is a strictly decreasing and strictly convex function of time: it decreases 
from $w(0)=\infty$ to $w(\infty)=0$. The Gauss hypergeometric functions ${_{2}F_{1} }$
in Eqs. (\ref{F_M}) and (\ref{P_M}) share the same parameters, but have distinct
time-dependent variables. We stress that, while the expression (\ref{F_M}) of the fidelity
of evolution is a new result, the purity (\ref{P_M}) of  a thermalized number state
was first written and analyzed in Ref. \cite{PT2000b}. However, in the sequel we extend
this previous investigation.

The purity of a dissipating $M$-photon Fock state takes a simpler form:
\begin{align}
& {\cal P}_M(t)=(1-\eta)^{2M} \, {_{2}F_{1} }\left(-M, -M; \,1; \, 
\left( \frac{\eta}{1-\eta} \right)^2 \, \right),                    \notag \\ 
& ( {\bar n}_{\rm R}=0 ).
\label{PMn=0}
\end{align}
Coming back to the general formula (\ref{P_M}), the time derivative of the purity 
has an expression analogous to Eq. (\ref{dotF_M}):
\begin{align}
& \dot{\cal P}_M(t)=-(2\gamma) {\cal P}_M(t) \left\{ M+(2M+1) \, \frac{ {\bar n}_{\rm R} \, \eta }
{1+2{\bar n}_{\rm T}(t) }     \right.         \notag \\ 
& \left. + \left[ -M+M^2 w(t) \,\frac{ {_{2}F_{1} }\left(-M+1, -M+1; \,2; \, w(t) \right) }
{ {_{2}F_{1} }\left(-M, -M; \,1; \, w(t) \right) } \right]      \right.         \notag \\ 
& \left. \times \frac{1}{1-\eta} \right\}.
\label{dotP_M}
\end{align}

Further, the evolved purity (\ref{P_M}) has the following short-time approximation:
\begin{align}
& {\cal P}_M(t)=1-2\left[ M+(2M+1) {\bar n}_{\rm R} \right] (\gamma t)                         \notag   \\  
& +\frac{1}{2} \left[ 6M^2 +2\left( 12 M^2+6M+1 \right) {\bar n}_{\rm R}         \right.     \notag   \\
& \left. +8\left( 3M^2+3M+1 \right) {\bar n}_{\rm R}^2 \right] (\gamma t)^2                   \notag   \\
& +{\rm O}\left[ (\gamma t)^3 \right],      \qquad          (\gamma t \ll 1),
\label{P_Mst}
\end{align}
From Eq. (\ref{P_Mst}) we get the rate of mixing of the initial Fock state, which is 
${\Gamma}_{\rm m}(M, {\bar n}_{\rm R} ):=-\dot{\cal P}_M(0)
=2\gamma \left[ M+(2M+1) {\bar n}_{\rm R} \right] $.

It is important to write the large-time expansion of the purity (\ref{P_M}): 
\begin{align}
& {\cal P}_M(t)= \frac{1}{1+2{\bar n}_{\rm R} } 
\left\{ 1- 2 \, \frac{ M-{\bar n}_{\rm R} }{1+2{\bar n}_{\rm R} }\, \eta      \right.      \notag  \\
& \left. +\frac{1}{2} \,\frac{2}{ ( 1+2{\bar n}_{\rm R} )^2 } 
\left[ 4( M-{\bar n}_{\rm R} )^2 -M(M+1) \right] {\eta}^2        \right.       \notag  \\
& \left. +{\rm O}\left( {\eta}^3 \right) \right\},     \qquad      ( \gamma t \gg 1).
\label{P_Mlt}
\end{align}
According to Eq. (\ref{P_Mlt}), the purity at thermal equilibrium
\begin{equation}
\lim_{t \to \infty} [{\cal  P}_M(t)]= \frac{1}{1+2{\bar n}_{\rm R} }
= {{\rm Tr}\left\{ \left[ \hat{\rho}_{\rm T}( {\bar n}_{\rm R} ) \right]^2 \right\}}
\label{PMas}
\end{equation}
is reached from below if ${\bar n}_{\rm R} \leqq M$, and from above if ${\bar n}_{\rm R} >M$.
As a consequence, there are some few regimes of mixing which are illustrated in Fig. 3 (b). 
In view of the initial slope of its graph displayed by Eq. (\ref{P_Mst}), the purity (\ref{P_M}) 
has at least a minimum when ${\bar n}_{\rm R} \leqq M$, while this is not necessarily true
when ${\bar n}_{\rm R} >M$ \cite{PT2000b}. 

At the weak non-classicality threshold $t_{\rm w}$, Eq. (\ref{WM>0}), the variable (\ref{w(t)}) 
becomes equal to one: 
$$ w(t)=1 \; \iff  \; t= t_{\rm w}. $$
Therefore, at the time $t=t_{\rm w}$, the purity (\ref{P_M}) and its time derivative (\ref{dotP_M}) 
are expressed more compactly via the Gauss summation formula (\ref{Gauss}): 
\begin{align}
& {\cal P}_M(t_{\rm w} )=2^{-2M} \binom{2M}{M} \, \frac{1+{\bar n}_{\rm R} }{1+2{\bar n}_{\rm R} };
\label{PMG}
\end{align}
\begin{equation}
{\dot{\cal P} }_M(t_{\rm w} )=-\gamma {\bar n}_{\rm R} {\cal P}_M(t_{\rm w} ) \leqq 0.     
\label{dotPMG}
\end{equation}
The purity (\ref{PMG}) decreases to zero when the initial number of photons $M$
increases indefinitely.
In addition, for a dissipating mode $( {\bar n}_{\rm R}=0),$ the transient purity (\ref{PMn=0}) 
has a minimum at the threshold of weak non-classicality $t_{\rm w}=\frac{1}{\gamma} \ln(2),$
since 
$$\ddot{\cal P}_M(t_{\rm w} )=2^{-2(M-1)} \binom{2(M-1)}{M-1} \,{\gamma}^{2} >0,
\quad     ( {\bar n}_{\rm R}=0).$$
 
Despite their resembling analytic expressions, the fidelity of damping (\ref{F_M}) 
and the evolving purity (\ref{P_M}) have different monotonicity properties, as shown in Fig. 3.
As a matter of fact, the possible existence of a minimum of the purity ${\cal P}(t)$
during thermalization, which is smaller than its steady-state limit, was proven long ago 
for two classes of input pure states, namely,  the even coherent states \cite{PT2000a}
and the Fock states \cite{PT2000b}. The sufficient condition for the existence 
of such a minimum is that the initial mean photon number in the mode exceeds 
or is at least equal to the corresponding mean occupancy of the reservoir: 
$\langle {\hat a}^{\dag} \hat a \rangle_0 \geqq \bar n_R$. In Appendix C we prove
that this condition is not necessary.

Figure 3a) shows the evolution of the fidelity (\ref{F_M}) for several initial $M$-photon states
at a fixed thermal mean occupancy ${\bar n}_{\rm R}=0.5$. The decrease of the fidelity 
of evolution from one to the thermal-equilibrium value (\ref{FMas}) is stronger for larger $M$. 
In Fig. 3 b) we illustrate both regimes of evolution of the purity (\ref{P_M}) 
conditioned by the relation between the thermal mean occupancy ${\bar n}_{\rm R}$ 
and a fixed input number of photons $M=2$: 

i) the non-monotonicity property of the evolving purity (\ref{P_M}), which exhibits a minimum 
when ${\bar n}_{\rm R} \leqq M;$ 

ii) the strict decrease of the purity (\ref{P_M}) under the reverse condition.

\begin{figure}[h]
\center
\includegraphics[width=9cm]{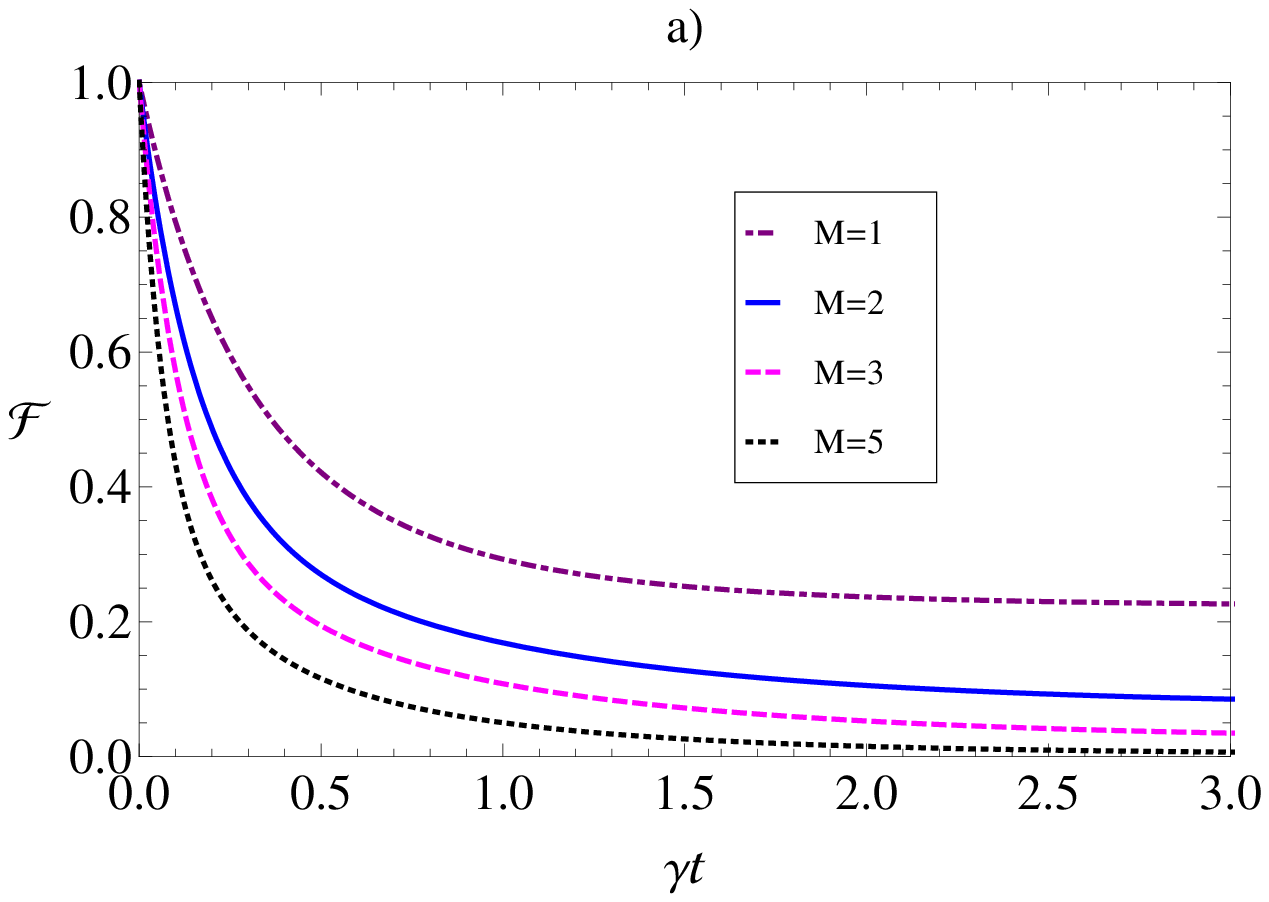}
\includegraphics[width=9cm]{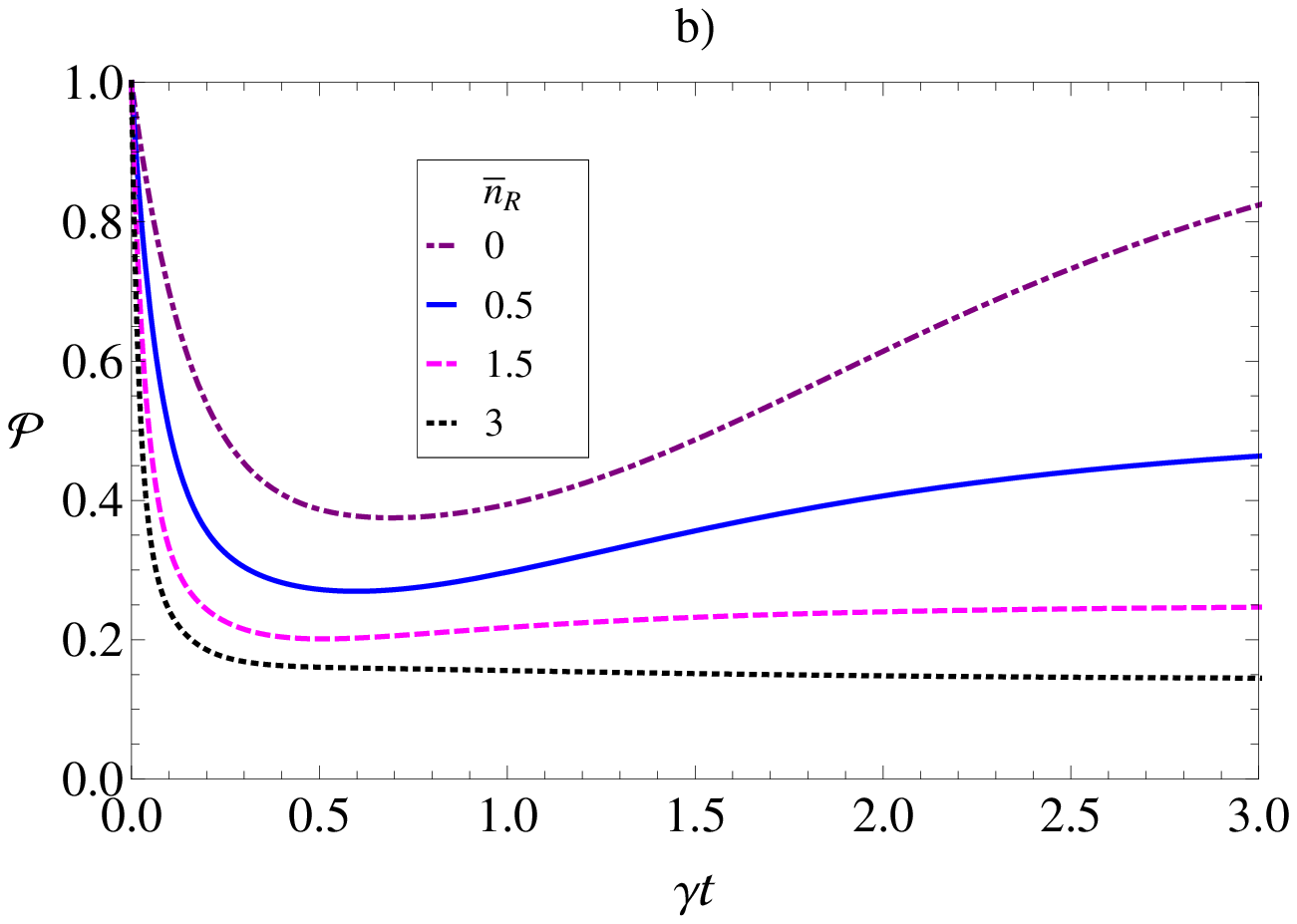}
\caption{(Color online) (a) The decrease of the fidelity of evolution ${\cal F}_M(t)$, 
Eq. (\ref{F_M}), is faster for larger $M$. We plot it for ${\bar n}_{\rm R}=0.5$ 
and the following initial numbers of photons: $M=1$ (dotted-dashed purple line), 
$M=2$ (solid blue line), $M=3$ (dashed magenta line), and $M=5$ (dotted black line).
(b) The evolution of the purity ${\cal P}_M(t)$, Eq. (\ref{P_M}), is not monotonic 
for $M \geqq {\bar n}_{\rm R}$.  At a fixed initial number of photons $M=2$, we present 
its plots for several values of the thermal mean occupancy: ${\bar n}_{\rm R}=0$
(dotted-dashed purple line), ${\bar n}_{\rm R}=0.5$ (solid blue line),
${\bar n}_{\rm R}=1.5$ (dashed magenta line), and ${\bar n}_{\rm R}=3$ (dotted black line).}
\end{figure}

We now consider the HS distance of evolution (\ref{HS1}) adapted to this section:
\begin{equation} 
{\cal G}_M(t)=\left[ 1+ {\cal P}_M(t)-2 {\cal F}_M(t) \right]^{\frac{1}{2} }.
\label{G_M}
\end{equation}
This is a feasible indicator of a Fock-state damping , being built with the fidelity 
of evolution (\ref{F_M}) and the purity (\ref{P_M}). Two remarks are  noteworthy. 
First,  Eqs. (\ref{F_Mst}) and (\ref{P_Mst}) provide the short-time behavior 
of the HS distance of evolution (\ref{G_M}): 
\begin{equation}
{\cal G}_M(t)=\dot{\cal G}_M(0) \, t +{\rm O}\left[ (\gamma t )^2 \right], 
\qquad          ( \gamma t \ll 1),
\label{G_Mst}
\end{equation}
with the slope at the origin
\begin{align}
\dot{\cal G}_M(0)=\sqrt{2} \gamma \left[ \left( M-{\bar n}_{\rm R} \right)^2 
+ 3M(M+1){\bar n}_{\rm R}(1+{\bar n}_{\rm R}) \right]^{ \frac{1}{2} }.
\label{dotG_M(0)}
\end{align}
Second, from Eqs. (\ref{F_Mlt}) and (\ref{P_Mlt}), we infer that the steady-state limit
\begin{align}
\lim_{t \to \infty} [\, {\cal  G}_M(t)]=\sqrt{2}\left[ \frac{1+{\bar n}_{\rm R} }{1+2{\bar n}_{\rm R} }       
-\frac{1}{1+{\bar n}_{\rm R} } \left( \frac{ {\bar n}_{\rm R} }{1+{\bar n}_{\rm R} } \right)^M  
\right]^{ \frac{1}{2} }
\label{GMas}
\end{align}
is attained from below if ${\bar n}_{\rm R} \leqq M$.

Figure 4 displays the HS distance of evolution (\ref{G_M}) at a fixed thermal mean occupancy 
${\bar n}_{\rm R}=0.5$, for the same input Fock states as in Fig. 3 a). All four plots drawn here 
point out a convenient monotonic increase of the function ${\cal G}_M(t)$, i. e., 
of the HS distinguishability between the initial number state and the evolved one 
through the master equation (\ref{QOME}). A rapid increase towards its input-output value
(\ref{GMas}) shows that the interaction with the reservoir makes the states  ${\hat \rho}_M(0)$ 
and ${\hat \rho}_M(t)$ to become more and more distinguishable. This happens faster 
for larger $M$, confirming that the fragility of a Fock state under the influence 
of thermal noise increases with $M$ \cite{PT2000b}. 

\begin{figure}[h]
\center
\includegraphics[width=9cm]{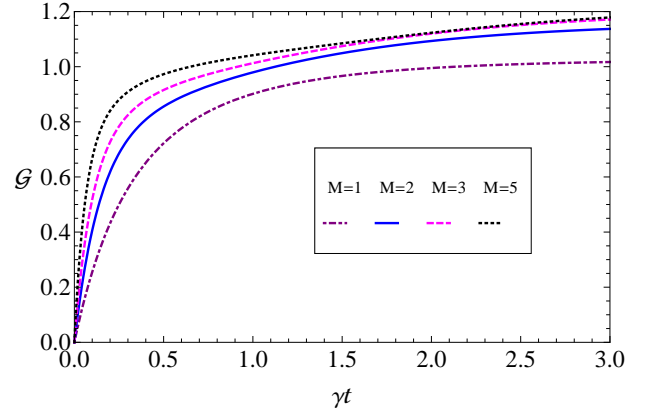}
\caption{(Color online) The monotonic behavior of the Hilbert-Schmidt distance of evolution
${\cal G}_M(t)$,  Eq. (\ref{G_M}), for ${\bar n}_{\rm R}=0.5$ and the same initial photon 
numbers $M$ as in Fig. 3 a).}
\end{figure}

\subsubsection{Upper bounds for the speed of dissipation}
 
By reason of feasibility, we restrict the next analytic treatment to the case of dissipation
$( {\bar n}_{\rm R}=0 )$ of a mode whose initial state has only one or two photons.
Then, Eq. (\ref{CFMt}) simplifies to:
\begin{equation}
\chi_M(\lambda, {\lambda}^{\ast}, t)=\exp\left( -\frac{1}{2} |\lambda|^2 \right)
L_M \left( \eta |\lambda|^2 \right).
\label{CFMdis}
\end{equation}
On account of Eqs. (\ref{eta}) and (\ref{1F1}), we get the general formula:
\begin{align}
\frac{\partial}{\partial t}\left[ {\chi}_M^{(N)}(\lambda, {\lambda}^{\ast}, t) \right]
=M \gamma \eta \, |\lambda|^2 {_{1}F_{1} }(-M+1; \, 2; \,\eta \, |\lambda|^2).
\label{dotCFMdis}
\end{align}
In particular,
\begin{equation}
\frac{\partial}{\partial t}\left[ {\chi}_1^{(N)}(\lambda, {\lambda}^{\ast}, t) \right]
=\gamma \eta \, |\lambda|^2;
\label{dotCF1dis}
\end{equation}
\begin{equation}
\frac{\partial}{\partial t}\left[ {\chi}_2^{(N)}(\lambda, {\lambda}^{\ast}, t) \right]
=2 \gamma \eta \, |\lambda|^2 \left( 1-\frac{1}{2} \eta \, |\lambda|^2 \right).
\label{dotCF2dis}
\end{equation}
Use of Eq. (\ref{v^2}) provides the corresponding upper bounds for the speed of dissipation:
\begin{align}
& {\tilde v}_1(t)=\sqrt{2} \gamma \eta,        \notag      \\
& {\tilde v}_2(t)=2\sqrt{6}\, \gamma \eta \left( {\eta}^2 -\eta +\frac{1}{3} \right)^{\frac{1}{2} }.
\label{tv1,2}
\end{align}

\subsection{Quantum speed limit times for dissipation}

We specialize  Eq. (\ref{v(M;0)}) to write the following constant QSLs for the dissipation 
of the field mode:
\begin{equation}
v_{\cal F}(1;0)=\sqrt{5} \, \gamma,      \qquad            v_{\cal F}(2;0)=\sqrt{13} \, \gamma.  
\label{vF1,2}
\end{equation}
Then, insertion of the speeds (\ref{tv1,2}) into Eq. (\ref{QSL2}) gives the time-dependent 
QSLs for dissipation $({\bar n}_{\rm R}=0)$:
\begin{equation}
\overline{ {\tilde v} }_1(t)=\frac{\sqrt{2} }{t} (1-\eta), 
\label{btv_1}
\end{equation}
and, respectively,
\begin{align}
& \overline{ {\tilde v} }_2(t)=\frac{1}{t} \frac{\sqrt{6} }{12} \left\{ 2\sqrt{3}
-\ln\left[ \left( \eta -\frac{1}{2} \right) +\left( {\eta}^2 -\eta +\frac{1}{3} \right)^{\frac{1}{2} } \right]    
\right.     \notag   \\   
& \left. +\ln\left( \frac{1}{2}+\frac{1}{\sqrt{3} } \right)
-12\left( \eta -\frac{1}{2} \right) \left( {\eta}^2 -\eta +\frac{1}{3} \right)^{\frac{1}{2} }  \right\}.
\label{btv_2}
\end{align}

From Eq. (\ref{PMn=0}) one gets
\begin{equation}
{\cal P}_1(t)=(1-\eta)^2 +{\eta}^2 ;
\label{P1n=0}
\end{equation}
\begin{equation}
{\cal P}_2(t)=(1-2\eta)^2 + 6\,{\eta}^2 (1-\eta)^2.
\label{P2n=0}
\end{equation}
Taking account of the fidelity (\ref{F_Mn=0}) and the purities (\ref{P1n=0}), (\ref{P2n=0}),
we get the corresponding values of the HS evolution distance (\ref{G_M}):
\begin{equation} 
{\cal G}_1(t)=\sqrt{2}(1-\eta);
\label{G_11}
\end{equation}
\begin{equation} 
{\cal G}_2(t)=\sqrt{2}(1-\eta) \left( 1+3{\eta}^2 \right)^{\frac{1}{2} }.
\label{G_2}
\end{equation}

We now are ready to evaluate the fidelity QSLTs (\ref{tauFmin}),
\begin{equation}
{\tau}_{\cal F}^{\rm min}(M;t)=\frac{1-{\cal F}_M(t) }{v_{\cal F}(M;0) },
\qquad             (M=1, \, 2),
\label{tauFM}
\end{equation}
and (\ref{ttauF}),
\begin{equation}
\tilde{\tau}_{\cal F}(M;t)=\frac{1-{\cal F}_M(t) }{\overline{ {\tilde v} }_M(t)},
\qquad             (M=1, \, 2),
\label{ttauFM}
\end{equation}
as well as the HS-metric ones (\ref{ttauG}),
\begin{equation}
\tilde{\tau}_{\cal G}(M;t)=\frac{ {\cal G}_M(t) }{\overline{ {\tilde v} }_M(t)},
\qquad             (M=1, \, 2).
\label{ttauGM}
\end{equation}
With appropriate substitutions in Eqs. (\ref{tauFM})-(\ref{ttauGM}),
we get two pairs of fidelity-based QSLTs:
\begin{equation} 
{\tau}_{\cal F}^{\rm min}(1; t)=\frac{1-\eta}{\sqrt{5} \, \gamma },         \qquad
{\tau}_{\cal F}^{\rm min}(2; t)=\frac{1-{\eta}^2}{\sqrt{13} \, \gamma };
\label{tauF1,2}
\end{equation}
\begin{equation} 
\tilde{\tau}_{\cal F}(1; t)=\frac{\sqrt{2} }{2} t,                \qquad
\tilde{\tau}_{\cal F}(2; t)=\frac{1-{\eta}^2}{\overline{ {\tilde v} }_2(t) };
\label{ttauF1,2}
\end{equation}
and a pair of HS-metric-based QSLTs:
\begin{equation} 
\tilde{\tau}_{\cal G}(1; t)=t ,           \quad
\tilde{\tau}_{\cal G} (2; t)=\frac{\sqrt{2} }{\overline{ {\tilde v} }_2(t) }
(1-\eta) \left( 1+3{\eta}^2 \right)^{\frac{1}{2} }.
\label{ttauG1,2}
\end{equation}
The HS-metric QSLTs (\ref{ttauG1,2}) are obviously tighter than the analogous 
fidelity QSLTs (\ref{ttauF1,2}), in agreement with the general inequality (\ref{ttauGF}). 
Moreover, the time bounding (\ref{ttauG}) is saturated for  $M=1$.

Figure 5 illustrates the evolution of the above QSLs and their associate QSLTs 
during the dissipation of one- and two-photon states, which are of primary 
experimental interest.
\begin{figure}[h]
\center
\includegraphics[width=9cm]{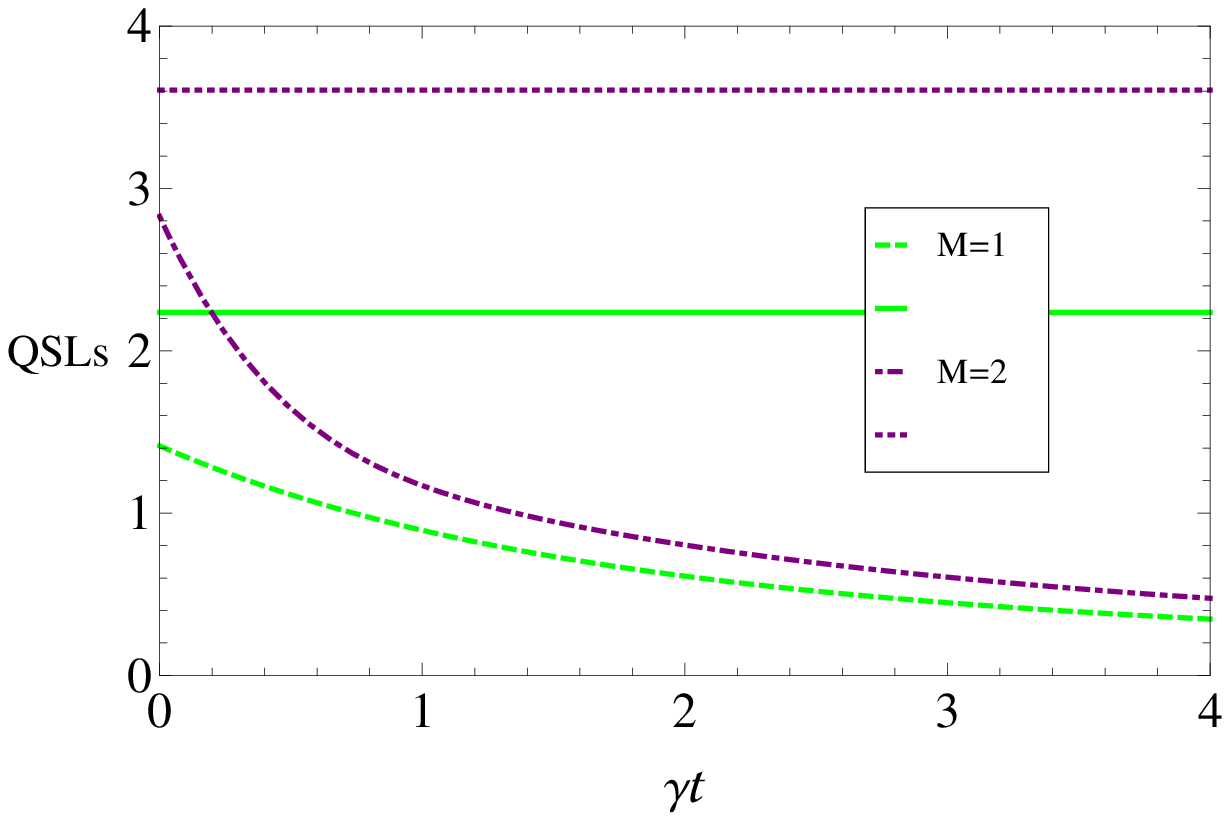}
\includegraphics[width=9cm]{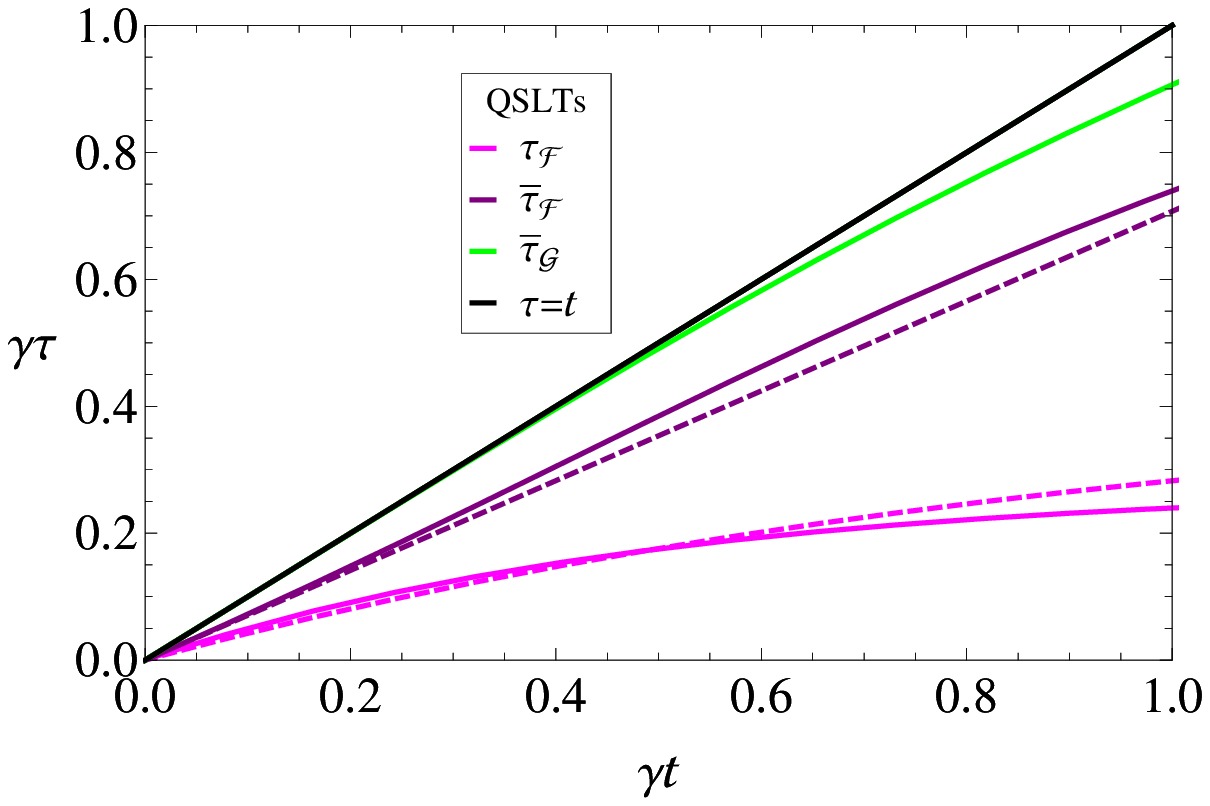}
\caption{(Color online) (up) The static QSLs of dissipation, Eq. (\ref{vF1,2}), for $M=1$ 
(horizontal solid green line) and $M=2$ (horizontal dotted purple line), together 
with the variable ones, Eq. (\ref{btv_1}), for $M=1$ (dashed green line), and Eq. (\ref{btv_2}),
for $M=2$ (dotted-dashed purple line). \\
(down) QSLTs for one- and two-photon input states of a mode weakly coupled 
to a zero-temperature reservoir. The fidelity-based QSLTs, Eq. (\ref{tauF1,2}), for $M=1$  
(bottom dashed magenta line) and $M=2$ (bottom solid magenta line), as well as
Eq. (\ref{ttauF1,2}), for $M=1$ (middle dashed purple line) and $M=2$ (middle solid purple line),
vs. those built with the Hilbert-Schmidt metric, Eq. (\ref{ttauG1,2}), for $M=1$ 
(top dashed green line superposed on the top solid black line) and $M=2$
(top solid green line). The tightness of these lower bounds to the actual time of interaction 
(top solid black line) is clearly visualized.} 
\end{figure}

\section{Summary and conclusions}

Quite long after the pioneering derivation of a more insightful time-energy uncertainty 
relation by Mandelstam and Tamm in 1945, many efforts were orientated towards
the evaluation of upper bounds on the speed of evolution for various types 
of  quantum dynamics: unitary evolution, open- and multipartite-system dynamics, 
as well as an extension to the evolution of mixed states \cite{GLM}.  We mention some
of the most significant advances on quantum evolution issues, which have been obtained 
so far \cite{ML,AA,V,Taddei,Campo,DL,Modi2,WuYu18,Pires,LZ,Campo1,Dod}. 

The present work is devoted to the evaluation of QSLTs for a well-known continuous-variable 
open system: a cavity field mode weakly coupled to a bosonic reservoir. The field is initially 
in a pure state and its evolution is a Markovian thermalization, being ruled by the quantum 
optical master equation, Eq. (\ref{QOME}). We develop here the ideas and exploit the tools 
put forward in the remarkable papers by del Campo, Egusquiza, Plenio, and Huelga 
\cite{Campo}, Deffner and Lutz \cite{DL}, and Campaioli, Pollock, and Modi \cite{Modi2}. 
In Refs. \cite{Campo,DL}, the QSL is built by using the fidelity of evolution (\ref{fidev}), while 
Ref. \cite{Modi2} employs instead the HS distance of evolution, Eqs. (\ref{HS})-(\ref{HS1}).

The QSL (\ref{vF0}) derived in Ref. \cite{Campo} is time-independent, being determined 
by the initial pure state of the quantum system, as well as by the unperturbed Hamiltonian 
and the Lindblad operators occurring in the master equation. The associate QSLT (\ref{tauFmin}),
besides its feasibility, is the natural generalization of that describing the pure-state evolution 
of an isolated quantum system. Here we have tightened the QSLT (\ref{tauFmin}) 
to another one, Eq. (\ref{tauF}), whose  time-dependent QSL (\ref{QSL1}) involves 
the time-averaged square root of the current purity, $\overline{\sqrt{ {\cal P} } }(t)$. Anyway, 
its evaluation requires the knowledge of the solution ${\hat \rho}(t)$ of the master equation.

More important, the common upper bound (\ref{tv}) found in Ref. \cite{DL} for the rate of
the fidelity of evolution and in Ref. \cite{Modi2} for the rate of the HS distance of evolution 
leads to the same time-dependent QSL (\ref{QSL2}). We have proven in a straightforward way
this equality, provided that the initial state is pure. According to the inequality (\ref{ttauGF}),
which follows immediately, the QSLT (\ref{ttauG}), built with the HS metric, is tighter than
the fidelity-based QSLT (\ref{ttauF}). However, in order to evaluate the QSL (\ref{QSL2}),
one needs to solve the master equation and then to handle its solution analytically or
numerically.

We derive an expression of the time-independent QSL $v_{\cal F}(0)$, Eq. (\ref{vF0}), 
valid for any input pure one-mode state. This general formula is then applied to a couple of 
complete systems of states which are widely used in quantum physics: the coherent states 
and the Fock states. In order to evaluate the other two time-dependent QSLs, we recall
the CF of the evolved state under damping. Then we use it to write, via the Weyl expansion,  
the general expressions of three main ingredients: the fidelity of evolution, the evolved purity, 
and the upper bound for the speed of evolution.

In the case of coherent states, in order to use the fidelity of evolution as a figure of merit,
we have to replace its expression in the Schr\"odinger picture, Eq. (\ref{Fcoh}), which is 
an oscillating function, by its interaction-picture counterpart, Eq. (\ref{F1}). This one
is a convenient approximate fidelity, strictly decreasing in time, and thereby it provides 
the desired good distinguishability of the evolved states.

The exact expression of the upper bound for the squared speed of evolution (\ref{vcoh^2})
has two terms. The former is an excellent approximation in the optical domain, 
while the latter describes quite accurately the damping of a low-intensity laser radiation. 
Besides the exact formula (\ref{v(t)coh}) established for the QSL $\overline{v_{\cal F} }(t)$,
we derive and analyze the QSLs $\overline{\tilde v}(t)$, Eqs. (\ref{btv1}) and (\ref{btv2}), 
obtained with the above-mentioned approximations. The associate QSLTs of the former
become exact for dissipation, but are not tight; the loosest bound is that corresponding
to the QSL $v_{\cal F}(0)$. A separate discussion is devoted to the thermalization
of the vacuum state, where all results are exact.

Then we review the evolution of any excited Fock state during thermalization \cite{PIT13}.
This process consists of three successive stages: strong non-classicality, weak non-classicality,  
and classicality. They are separated by two threshold times, $t_{\rm w}$ and $t_{\rm c}$, 
which are determined by the equilibrium photon number ${\bar n}_{\rm R}$, 
being independent of the initial number of photons $M$. The exact formulas established
for the fidelity of evolution, Eq. (\ref{F_M}), the evolving purity, Eq. (\ref{P_M}), and
the HS distance of evolution, Eq. (\ref{G_M}), valid for any $M$, are thoroughly 
analyzed. Here we content ourselves with evaluating the QSLs (\ref{vF1,2})-(\ref{btv_2}) 
for dissipation of one- and two-photon input states. By examining the associate QSLTs, 
Eqs. (\ref{tauF1,2})-(\ref{ttauG1,2}), we remark the quite impressive accuracy of those 
based on the HS distance of evolution. Moreover, according to Eq. (\ref{ttauG1,2}), 
the inequality (\ref{ttauG}) is even saturated for $M=1$. In Appendix C, a special 
attention is paid 
to the damping of a one-photon state. We focus on a conditioned non-monotonic 
evolution of the HS distance ${\cal G}_1(t)$, Eq.(\ref{G_1}), essentially due to a similar 
behavior of the purity ${\cal P}_1(t)$.  Nevertheless, concerning the damping of other  
Fock states, a numerical approach is at hand for all needed cases. 

We are left to draw some noteworthy conclusions. 
\begin{itemize}
\item{The fidelity-based QSL $v_{\cal F}(0)$, Eq. (\ref{vF0}), derived in Ref. \cite{Campo}, 
is rather unique, owing to the remarkable feature of being time-independent. This makes it 
possible to get out of solving the master equation that describes the Markovian dynamics
of an open system. As far as we know, Eq. (\ref{v(0)}) is its first application 
to a continuous-variable setting.}
\item{The time-dependent QSLs based on the fidelity of evolution and on the HS distance
of evolution are especially useful for non-Gaussian input states, when other fidelity 
approaches \cite{Taddei} are  analytically difficult.}
\item{For any initial pure state, our proof of the inequality (\ref{ttauGF}) is straightforward 
and general.}
\item{The QSLTs built using the HS metric, Eq. (\ref{ttauG}), have some attractive qualities: 
they are feasible, tight, and robust under composition \cite{Modi2}. For instance, 
the dissipation of one- and two-photon states displays a surprisingly good tightness 
of the QSLTs based on the HS metric.}
\item{Even in the simplest case of the damping of a one-photon state, 
the HS distance of evolution ${\cal G}_1(t)$, Eq.(\ref{G_1}), could exhibit a slightly 
non-monotonic behavior. However, these possible deviations from monotonicity 
are too small to alter its value as a suitable QSL.}
\end{itemize}

\appendix

\section{Smoothing the fidelity of damping of a coherent state}

In order to smooth the fidelity of evolution (\ref{Fcoh}), one should replace 
the oscillating function $\cos(\omega t)$ with one or another of its most significant values.
We choose here its extrema, as well as its zero value, to get the approximate fidelities:
\begin{align} 
{\cal F}^{(\pm 1)}(t)=\frac{1}{ 1+{\bar n}_{\rm T}(t) } 
\exp \left[ -|\alpha|^2 \, \frac{ \left( 1\mp \sqrt{\eta} \right)^2 }{ 1+{\bar n}_{\rm T}(t) } \right],      
\label{F1,-1}
\end{align} 
\begin{align}
{\cal F}^{(0)}(t)& =\frac{1}{ 1+{\bar n}_{\rm T}(t) } \exp \left[ - |\alpha|^2
\, \frac{1+\eta} { 1+{\bar n}_{\rm T}(t) } \right]       \notag   \\    
& = \left[ {\cal F}^{(1)}(t) \, {\cal F}^{(-1)}(t) \right]^{\frac{1}{2} }.
\label{F0}
\end{align} 
We also contemplate two other reasonable averages. The former is the arithmetic mean 
of the smoothed fidelities (\ref{F1,-1}):
\begin{align}
{\cal F}^{(A)}(t): & = \frac{1}{2} \left[ {\cal F}^{(1)}(t)
+{\cal F}^{(-1)}(t) \right]            \notag   \\    
& = {\cal F}^{(0)}(t) \cosh\left[ \frac{ 2|\alpha|^2 \, \sqrt{ \eta} }{ 1+{\bar n}_{\rm T}(t) } \right].
\label{FA}
\end{align} 
The latter is the time average over a period $T={2\pi}/{\omega}$ of the fidelity
of evolution (\ref{Fcoh}), which is considered to be a function of the variable $\omega t$ only,
with the large-scale parameter (\ref{eta}) kept constant: 
\begin{align}
{\cal F}^{(T)}(t): & =\frac{1}{2\pi} \int_{-\pi}^{\pi} d(\omega t) \, {\cal F}(t)        
\notag   \\    
& ={\cal F}^{(0)}(t) \, I_0 \left[ \frac{ 2|\alpha|^2 \, \sqrt{ \eta} }{ 1+ {\bar n}_{\rm T}(t) } \right].
\label{FT}
\end{align} 
In Eq(\ref{FT}),  $I_0(z)$ is the modified Bessel function of the first kind and order zero \cite{HTF2}:
\begin{equation}
I_0(z)=\sum_{m=0}^{\infty} \frac{1}{m! \, m!}\left( \frac{1}{2}z \right)^{2m},     \qquad
(z \in {\mathbb C}).
\label{I_0}
\end{equation}

The inequalities $ {\cal F}^{(1)}(t) \geqq  {\cal F}(t)  \geqq  {\cal F}^{(-1)}(t)$
become saturated at times $$\, \omega t= 2n\pi    \; \; \;  \text{and} \; \; \;
\omega t=(2n+1)\pi,      \quad    (n=0, 1, 2, 3, ...), $$ respectively. They show 
that the graphs of the functions (\ref{F1,-1}) are upper and lower envelopes 
for the oscillating fidelity of evolution (\ref{Fcoh}), which are periodically touched. 
On the other hand, from the general inequalities
$${\rm e}^x > \cosh(x) > I_0(x) >1> {\rm e}^{-x} >0,    \qquad     (x>0), $$
we infer the following strict hierarchy, valid for any finite time $t \geqq 0$,
provided that $|\alpha| >0$:
$${\cal F}^{(1)}(t) > {\cal F}^{(A)}(t) > {\cal F}^{(T)}(t) > {\cal F}^{(0)}(t) > {\cal F}^{(-1)}(t) >0.$$
However, all five smoothed approximate fidelities (\ref{F1,-1})-(\ref{FT}) share 
the same steady-state limit (\ref{FcohIO}), which is precisely that of the exact fidelity 
of evolution (\ref{Fcoh}).

Figure 6 illustrates the above-mentioned hierarchy of smoothed approximate fidelities 
of evolution (\ref{F1,-1})-(\ref{FT}) superposed on the oscillating exact one, Eq. (\ref{Fcoh}).

\begin{figure}[h]
\center
\includegraphics[width=9cm]{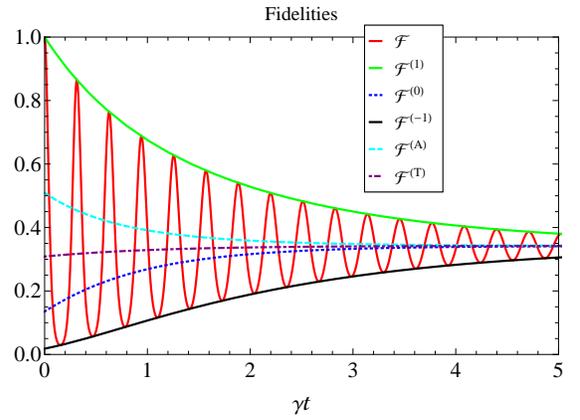}
\caption{(Color online) The smoothed approximate fidelities (\ref{F1,-1})-(\ref{FT}) 
vs. the oscillating exact fidelity of evolution (\ref{Fcoh}) for $|\alpha |^2=2$: 
${\cal F}(t)$ (oscillating solid red curve), ${\cal F}^{(1)}(t)$ (upper solid green envelope), 
${\cal F}^{(-1)}(t)$ (lower solid black envelope), ${\cal F}^{(0)}(t)$ (dotted blue line), 
${\cal F}^{(A)}(t)$ (dashed cyan line), ${\cal F}^{(T)}(t)$ (dotted-dashed purple line). 
The ratio $\omega /\gamma=10$ chosen here is much smaller than the value 
$\omega/ \gamma \approx 10^6$ met in an optical cavity.}
\end{figure}

Let us introduce the function ${\cal F}^{(1)}(t)$ instead of ${\cal F}(t)$ into Eq. (\ref{HS1}), 
in order to replace the HS evolution distance ${\cal G}(t)$ by its modified version,
\begin{equation} 
{\cal G}^{(1)}(t):=\left[ 1+ {\cal P}(t)-2{\cal F}^{(1)}(t) \right]^{\frac{1}{2} }.
\label{GC1}
\end{equation}
Among the five selected candidates (\ref{F1,-1})-(\ref{FT}), the function ${\cal F}^{(1)}(t)$
is the only one to possess the following two properties, valid at any time and for all values
of their parameters $|\alpha|$ and ${\bar n}_{\rm R}$:

1) Its associate figures of merit fulfil the inequalities 
$$ 1-{\cal F}^{(1)}(t) \leqq 1-{\cal F}(t), \qquad   {\cal G}^{(1)}(t) \leqq {\cal G}(t),$$
which are necessary for preserving the QSLs  $\overline{ v_{\cal F} }(t)$,  Eq. (\ref{QSL1}), 
and, respectively,  $\overline{ \tilde{v} }(t)$,  Eq (\ref{QSL2}), 
when we replace the fidelity ${\cal F}(t)$ by ${\cal F}^{(1)}(t)$;

2) The function ${\cal F}^{(1)}(t)$ is strictly decreasing, regardless of the values
taken by its parameters $|\alpha|$ and ${\bar n}_{\rm R}$: it decreases from the initial value 
${\cal F}^{(1)}(0)=1$ to reach from above the asymptotic limit (\ref{FcohIO}).
This behavior is particularly attractive because it endows the approximate fidelity 
${\cal F}^{(1)}(t)$ with the best distinguishability between evolved states, 
and thereby secures a maximal efficiency of the above-mentioned figures of merit.

Moreover, ${\cal F}^{(1)}(t)$ is the fidelity of evolution in the interaction picture,
\begin{equation}
{\cal F}^{(1)}(t)=\langle \alpha |\left( {\hat \rho}_{\, \rm DT} \right)_I(t)| \alpha \rangle,
\label{FcohIP}
\end{equation}
which is the smooth version of its counterpart in the Schr\"odinger picture, Eq. (\ref{FcohSP}).
Indeed, in Eq. (\ref{FcohIP}), the state $ \left( {\hat \rho}_{\, \rm DT} \right)_I(t)$ 
stands for the transform in the interaction picture of the DTS (\ref{DTS}),
which is given in the Schr\"odinger picture \cite{hybridCF}. In general, being a two-time 
transition probability, such a fidelity of evolution is picture-dependent.

To sum up, the only smooth approximate fidelity of evolution that can replace 
the oscillating exact one ${\cal F}(t)$, Eq. (\ref{Fcoh}), without any damage,
is its interaction-picture counterpart ${\cal F}^{(1)}(t)$, Eq. (\ref{F1}). For $\alpha \neq 0,$ 
this will consistently be done in the present paper.

\section{An infinite integral involving Laguerre polynomials}

Our aim here is to evaluate the integral
\begin{align}
& {\mathcal L}_M (s, \sigma, {\sigma}^{\prime}):=\int_0^{\infty} dt \exp(-st) L_M(\sigma t)
L_M( {\sigma}^{\prime} t),         \notag   \\
& (s \geqq \sigma >0, \;\;s \geqq {\sigma}^{\prime} >0 ).
\label{LTLL}
\end{align}
Throughout this paper we denote the Laguerre polynomial of degree $M$ by $L_M(x)$.
To write it explicitly, it is convenient to use its expression as a confluent hypergeometric
function ${_{1}F_{1}}$ \cite{HTF2a}:
\begin{align}
L_M(z)={_{1}F_{1}}(-M; 1; z)=\sum_{m=0}^{\infty} \frac{ (-M)_m }{ (1)_m \, m! }\, z^m , \label{1F1}\end{align}   
with the Pochhammer symbol defined as follows:                              
\begin{align*}    &(a)_0:=1,  ( a)_m :=\frac{ \Gamma(a+m) }{ \Gamma(a) }=a(a+1)...(a+m-1), \nonumber\\&                                
 (m=1,2,3,...).\end{align*}
We exploit a particular case of the Hille-Hardy summation formula \cite{HTF2a}:
\begin{align}
& \sum_{m=0}^{\infty} L_m(x)  L_m(y) z^m      \notag  \\   
& = (1-z)^{-1} \exp\left[ -\frac{z}{1-z} (x+y) \right] I_0\left[ 2\frac{ (xyz)^{\frac{1}{2} } }{1-z} \right],   
\notag  \\  
& (|z| <1),
\label{HH}
\end{align}
where $I_0(z)$ is the modified Bessel function of the first kind and order zero, Eq. (\ref{I_0}).
The function on the r. h. s. of Eq. (\ref{HH}) is called the bilinear generating function 
of the Laguerre polynomials  (\ref{1F1}).
One gets the Laplace transform of the function $I_0(bt)$ by integrating its Maclaurin series
(\ref{I_0}) term by term:
\begin{align}
& \int_0^{\infty} dt \exp(-at) I_0(bt)=\left( a^2-b^2 \right)^{-\frac{1}{2} },          \notag  \\    
& (a>b \geqq 0).
\label{LTI_0}
\end{align}
We choose $x= {\sigma}t, \, y= {\sigma}^{\prime}t, \, z \in (0,1) \subset {\mathbb R}_{+}$
in Eq. (\ref{HH}) and then introduce the positive parameters
\begin{align}
a=s+\frac{z}{1-z} ( {\sigma}+{\sigma}^{\prime} ) \text{\;\; and \;\; }
b=2\frac{z^{\frac{1}{2} } }{1-z} ( {\sigma} {\sigma}^{\prime} )^{\frac{1}{2} }
\label{a,b}
\end{align}
into Eq. (\ref{LTI_0}).  As a result, Eqs. (\ref{LTLL}), (\ref{HH}), and (\ref{LTI_0})
yield the sum of the following power series:
\begin{align}
\sum_{m=0}^{\infty} {\mathcal L}_m (s, \sigma, {\sigma}^{\prime}) z^m
=\left[ (1-z)a \right]^{-1}\left( 1-\frac{b^2}{a^2} \right)^{-\frac{1}{2} },
\label{sum}
\end{align}
provided that the condition $a^2-b^2>0$ is fulfilled. This is equivalent 
to the positivity of the quadratic trinomial
\begin{align}
& p(z):=Az^2-2Bz +C:   \qquad      A=(s-{\sigma}-{\sigma}^{\prime} )^2 \geqq 0,    \notag  \\ 
& B=(s-\sigma )(s-{\sigma}^{\prime} )+{\sigma}{\sigma}^{\prime}>0,    \qquad        C=s^2>0.     
\label{trinom}
\end{align}
One has the following alternative:

a) $A=0 \, \Longleftrightarrow \, s={\sigma}+{\sigma}^{\prime}.$ Then,
$$ p(z)=4{\sigma}{\sigma}^{\prime}(1-z)+({\sigma}{\sigma}^{\prime})^2>0, 
\qquad( z \leqq 1). $$

b) $A>0 \, \Longleftrightarrow \, s \neq {\sigma}+{\sigma}^{\prime}.$
Owing to its non-negative discriminant, 
$\, \Delta =16 \, {\sigma}{\sigma}^{\prime}(s-{\sigma})(s-{\sigma}^{\prime}) \geqq 0, $
the quadratic trinomial (\ref{trinom}) has two real roots that are both positive:
$$ z_{1,2}=\frac{1}{A}\left( B \mp \frac{1}{2} \sqrt{\Delta} \right) >0.$$
Accordingly, $p(z) >0 \;\; \text{for} \;\;  z<{\rm min}\{1,\, z_1\}.$

To sum up, there is always a subinterval of $ (0,1) \subset {\mathbb R}_{+}$ 
on which $\, p(z)>0 \,$ and therefore Eq. (\ref{sum}) is valid.

Taking into account the notations (\ref{a,b}), we expand the r.h.s. of Eq. (\ref{LTLL})
into a double binomial series and then rearrange it as an ascending power series
of the variable $z$, which converges in the interval specified above:
\begin{align}
& \left[ (1-z)a \right]^{-1}\left( 1-\frac{b^2}{a^2} \right)^{-\frac{1}{2} }
=\frac{1}{s} \sum_{m=0}^{\infty} \left[ \frac{s-( \sigma +{\sigma}^{\prime}) }{s} \right]^m 
z^m                                \notag  \\ 
& \times {_{2}F_{1}} \left (-m, m+1;1; - \frac{ \sigma {\sigma}^{\prime} }
{ s[s-( \sigma +{\sigma}^{\prime})] }  \right).
\label{sum1}
\end{align}
Recall that the Gauss hypergeometric function ${_{2}F_{1}}$ is the sum of a hypergeometric
series \cite{HTF1}:
\begin{align}
& {_{2}F_{1}}(a, b; \, c; \, z)=\sum_{m=0}^{\infty} \frac{ (a)_m (b)_m }{ (c)_m \, m! }\, z^m ,    
\notag  \\   
&  (c \neq 0, -1,-2,-3,...,        \qquad     |z| <1).                                                                          
\label{2F1}
\end{align}
Substitution of Eq. (\ref{sum1}) into Eq. (\ref{sum}) gives the formula:
\begin{align}
& {\mathcal L}_M (s, \sigma, {\sigma}^{\prime})
= \frac{1}{s} \left[ \frac{s-( \sigma +{\sigma}^{\prime}) }{s} \right]^M                \notag  \\ 
& \times {_{2}F_{1}} \left (-M, M+1; \,1;  \,- \frac{ \sigma {\sigma}^{\prime} }
{ s[s-( \sigma +{\sigma}^{\prime})] }  \right),
\label{LTLL1}
\end{align}
where the Gauss hypergeometric function is a polynomial of degree $M$. One gets 
an alternative expression by performing Pfaff's linear transformation \cite{HTF1a}, 
\begin{align}
{_{2}F_{1}}(a, b; \, c; \, z)=(1-z)^{-a} \, {_{2}F_{1}}\left( a, c-b; \, c; \, \frac{z}{z-1} \right),    \notag  
\end{align}
in Eq. (\ref{LTLL1}): 
\begin{align}
& {\mathcal L}_M (s, \sigma, {\sigma}^{\prime})
=  \frac{ (s- \sigma )^M (s-{\sigma}^{\prime})^M }{s^{2M+1} }                     \notag  \\ 
& \times {_{2}F_{1}} \left (-M, -M; \, 1; \,  \frac{ \sigma {\sigma}^{\prime} }
{ (s- \sigma) (s-{\sigma}^{\prime} ) }  \right),                                                \notag  \\ 
& (s \geqq \sigma >0, \;\;s \geqq {\sigma}^{\prime} >0 ).                            
\label{LTLL2}
\end{align}
In the special case $s=\sigma +{\sigma}^{\prime}$, by virtue of Gauss's summation theorem
\cite{HTF1b}, 
\begin{align}
& {_{2}F_{1}}(a, b; \, c; \, 1)=\frac{ \Gamma (c) \,\Gamma (c-a-b) }{ \Gamma (c-a) \, \Gamma (c-b) },
\notag   \\ 
& (c \neq 0, -1,-2,-3,...,        \quad       \Re(c-a-b)>0 ),  
\label{Gauss}
\end{align}
the polynomial (\ref{LTLL2}) can be written as a single monomial:
\begin{align} 
{\mathcal L}_M ( \sigma +{\sigma}^{\prime}, {\sigma}, {\sigma}^{\prime})
=\binom{2M}{M} \, \frac{( {\sigma}{\sigma}^{\prime} )^M }
{ (\sigma +{\sigma}^{\prime} )^{2M+1} }.        
\notag 
\end{align}

\section{Damping of a one-photon state}

The fidelity of evolution of an initial one-photon state is a strictly decreasing function of time, 
no matter how large is the mean photon occupancy ${\bar n}_R$ at thermal equilibrium. 
Indeed, for $M=1$, Eq. (\ref{dotF_M}) has a simpler form, whose sign is manifest:
\begin{align} 
& \dot{\cal F}_1(t)=-\frac{\gamma \eta}{ \left[ 1+{\bar n}_{\rm T}(t) \right]^4 }
\left( a{\eta}^2 -2b \eta +c \right) :                                                \notag     \\ 
&  a={\bar n}_{\rm R}^2 (1+ {\bar n}_{\rm R} ),                            \qquad                    
b={\bar n}_{\rm R} \left( {\bar n}_{\rm R}^2  -2 \right),                  \notag     \\ 
& c=( 1-{\bar n}_{\rm R} ) \left( 1-{\bar n}_{\rm R}^2  \right):        \qquad
\dot{\cal F}_1(t) <0.
\label{dotF_1}
\end{align}
The fidelity of evolution ${\cal F}_1(t)$ is plotted in Fig. 3 a).

Further, the purity rate (\ref{dotP_M}) reduces to:
\begin{align}
& \dot{\cal P}_1(t)=-\frac{ 2\gamma \eta }{ \left[ 1+2{\bar n}_{\rm T}(t) \right]^4 }     
\left( A{\eta}^2 +2B\eta +C \right) :              \notag    \\ 
&    A=2{\bar n}_{\rm R}\left[ 1+2{\bar n}_{\rm R} (1+ {\bar n}_{\rm R} ) \right],      \quad           
B=(1+2{\bar n}_{\rm R} )\left( 1-2{\bar n}_{\rm R}^2  \right),         \notag    \\ 
& C=-(1-{\bar n}_{\rm R} )(1+2{\bar n}_{\rm R} )^2.       
\label{dotP_1}
\end{align}
The roots of the quadratic trinomial in Eq. (\ref{dotP_1}),
\begin{align}
& {\eta}_{1,2}=\frac{1+2{\bar n}_{\rm R} }{2{\bar n}_{\rm R} \left[ 1+2{\bar n}_{\rm R} 
(1+ {\bar n}_{\rm R} ) \right] } \left\{ \left( 2 {\bar n}_{\rm R}^2 -1 \right)     \right.     \notag    \\ 
& \left. \pm  \left[ 1+2{\bar n}_{\rm R} (1- {\bar n}_{\rm R} ) \right]^{\frac{1}{2} } \right\}, 
\label{eta1,2}
\end{align}
allow us to distinguish the following situations:
\begin{itemize}
\item{ $\; {\bar n}_{\rm R}=0: \qquad  \qquad  {\eta}_1 =\frac{1}{2}, \;\;\;   {\eta}_2 =-\infty ;$ }
\item{ $\; 0< {\bar n}_{\rm R} <1: \qquad  {\eta}_1 \in (0,1),\;\;\;   {\eta}_2 <0 ;$ }
\item{ $\; {\bar n}_{\rm R}=1: \qquad  \qquad  {\eta}_1 =\frac{3}{5}, \;\;\;   {\eta}_2 =0 ;$ }
\item{ $\; 1< {\bar n}_{\rm R} <\frac{1}{2}\left( 1+\sqrt{3} \right): \qquad  0< {\eta}_2 <{\eta}_1 <1 ;$ }
\item{ $\; {\bar n}_{\rm R} =\frac{1}{2}\left( 1+\sqrt{3} \right): \qquad   \qquad   
{\eta}_2 ={\eta}_1=\frac{1}{2} ;$ }
\item{ $\; {\bar n}_{\rm R} >\frac{1}{2}\left( 1+\sqrt{3} \right): \qquad  {\eta}_2 ={\eta}_1^{\ast}, \;\;\;
\Im( {\eta}_1) >0 . $}
\end{itemize}
Accordingly, when ${\bar n}_{\rm R} \in [\, 0,1]$, the purity ${\cal P}_1(t)$ decreases 
from the intial value equal to one to its single minimum at the time 
$t_1=-\frac{1}{\gamma} \ln( {\eta}_1)$ and then increases to reach asymptotically
the equilibrium value (\ref{PMas}). By contrast, when 
${\bar n}_{\rm R} \in \left(1, \frac{1}{2} ( 1+\sqrt{3} ) \right),$ the time $t_1$
of minimal purity is followed by the time $t_2=-\frac{1}{\gamma} \ln( {\eta}_2)$ where 
the purity ${\cal P}_1(t)$ has a maximum greater than the steady-state value (\ref{PMas}).
For ${\bar n}_{\rm R}=\frac{1}{2} ( 1+\sqrt{3} )$, the purity  ${\cal P}_1(t)$ becomes a decreasing  
function of time, because the two extrema merge at $t_1=t_2=\frac{1}{\gamma} \ln(2)$ 
into a stationary point of inflection whose ordinate is slightly greater 
than the asymptotic limit (\ref{PMas}). Finally, when ${\bar n}_{\rm R} >\frac{1}{2} (1+\sqrt{3} )$,
the function of time ${\cal P}_1(t)$ is strictly decreasing. To sum up, in the case $M=1$, 
there are three regimes of mixing determined by the mean thermal photon occupancy 
${\bar n}_{\rm R}$:
\begin{enumerate}
\item{$\; 0 \leqq {\bar n}_{\rm R} \leqq 1:$ The purity starts to decrease and reaches its minimum
at the time $t_1$; then it increases attaining from below an asymptotic limit (\ref{PMas}) 
lying in the interval $\left[ \frac{1}{3}, 1 \right]; $}
\item{$\; 1< {\bar n}_{\rm R} < (1+\sqrt{3} )/2 :$  The purity decreases up to its minimum at $t_1$,
then increases to a maximum at the subsequent time $t_2$, and finally it decreases again 
reaching from above an asymptotic limit (\ref{PMas}) belonging to the interval 
$\left( \frac{1}{2+\sqrt{3} }, \frac{1}{3} \right); $}
\item{$\; {\bar n}_{\rm R} \geqq (1+\sqrt{3} )/2 :$ The purity ${\cal P}_1(t)$ decreases starting 
from one to attain eventually from above the limit (\ref{PMas}) with a positive value
smaller than or at most equal to $ \frac{1}{2+\sqrt{3} }$.}
\end{enumerate}

Let us write explicitly the HS distance of evolution (\ref{G_M}) for $M=1$:
\begin{align}
{\cal G}_1(t)=\frac{ \sqrt{2} (1-\eta) \left\{ \, p[ {\bar n}_{\rm R}; 
{\bar n}_{\rm T}(t) ] \right\}^{ \frac{1}{2} } }{ \left\{ \left[ 1+{\bar n}_{\rm T}(t) \right]  
\left[ 1+2{\bar n}_{\rm T}(t) \right] \right\}^{ \frac{3}{2} } },
\label{G_1}
\end{align}
where $p[ {\bar n}_{\rm R}; {\bar n}_{\rm T}(t) ]$ is a polynomial in the variable 
${\bar n}_{\rm T}(t)$, Eq. (\ref{nT}), with non-negative coefficients depending
on the parameter ${\bar n}_{\rm R}$,
\begin{align}
& p[ {\bar n}_{\rm R}; {\bar n}_{\rm T}(t) ]=\sum_{j=0}^4 c_j ( {\bar n}_{\rm R} ) 
\, [ {\bar n}_{\rm T}(t) ]^j >0:             \notag   \\
& c_0( {\bar n}_{\rm R} )=1+4{\bar n}_{\rm R}+7{\bar n}_{\rm R}^2,         \notag   \\
& c_1( {\bar n}_{\rm R} )=3\left( 1+3{\bar n}_{\rm R}+8{\bar n}_{\rm R}^2 \right),  
\notag  \\
& c_2( {\bar n}_{\rm R} )= 3+{\bar n}_{\rm R}+27{\bar n}_{\rm R}^2,        \notag  \\
& c_3( {\bar n}_{\rm R} )=1-6{\bar n}_{\rm R}+12{\bar n}_{\rm R}^2,        \qquad  
 c_4( {\bar n}_{\rm R} )=4{\bar n}_{\rm R}^2.
\label{p}
\end{align}
The value at $\eta=0$ of the polynomial (\ref{p}),
\begin{equation}
p( {\bar n}_{\rm R}; {\bar n}_{\rm R} )=(1+{\bar n}_{\rm R})(1+2{\bar n}_{\rm R})^2
\left[ (1+ {\bar n}_{\rm R})^2 +{\bar n}_{\rm R}^3 \right],
\label{pas}
\end{equation}
yields the steady-state limit of the HS evolution distance (\ref{G_1}):
\begin{align} 
\lim_{t \to \infty} [\, {\cal  G}_1(t)]=\frac{ \sqrt{2} \left[ (1+{\bar n}_{\rm R} )^2 
+{\bar n}_{\rm R}^3 \right]^{\frac{1}{2} } }
{ (1+{\bar n}_{\rm R}) (1+2{\bar n}_{\rm R})^{\frac{1}{2} } },       
\label{G1as}
\end{align}
in agreement with the general formula (\ref{GMas}).

Making use of Eqs. (\ref{G_1}) and (\ref{p}), we find the rate of change 
of the HS evolution distance ${\cal G}_1(t)$:
\begin{align}
\dot{\cal G}_1(t)=\frac{ \gamma \, \eta \, q[ {\bar n}_{\rm R}; {\bar n}_{\rm T}(t) ] 
 \left\{ 2p[ {\bar n}_{\rm R}; {\bar n}_{\rm T}(t) ] \right\}^{-\frac{1}{2} } }
{ \left\{ \left[ 1+{\bar n}_{\rm T}(t) \right] \left[ 1+2{\bar n}_{\rm T}(t) \right] \right\}^{\frac{5}{2} } },
\label{dotG_1}
\end{align}
where $q[ {\bar n}_{\rm R}; {\bar n}_{\rm T}(t) ]$ is a polynomial in the variable 
${\bar n}_{\rm T}(t)$, whose coefficients are functions of the parameter ${\bar n}_{\rm R}$,
\begin{align}
& q[ {\bar n}_{\rm R}; {\bar n}_{\rm T}(t) ]=\sum_{k=0}^5 d_k ( {\bar n}_{\rm R} ) 
\, [ {\bar n}_{\rm T}(t) ]^k :             \notag   \\
& d_0( {\bar n}_{\rm R} )=2\left( 1+4{\bar n}_{\rm R}+7{\bar n}_{\rm R}^2 \right),      \notag  \\
& d_1( {\bar n}_{\rm R} )=3\left( 2+5{\bar n}_{\rm R}+17{\bar n}_{\rm R}^2 \right),    \notag  \\
& d_2( {\bar n}_{\rm R} )=4\left( 1-7{\bar n}_{\rm R}+13{\bar n}_{\rm R}^2 \right),     \notag  \\
& d_3( {\bar n}_{\rm R} )=-\left( 4+81{\bar n}_{\rm R}+3{\bar n}_{\rm R}^2 \right),     \notag  \\
& d_4( {\bar n}_{\rm R} )=-2\left( 3+20{\bar n}_{\rm R}+6{\bar n}_{\rm R}^2 \right),   \notag  \\
& d_5( {\bar n}_{\rm R} )=2\left( -1+6{\bar n}_{\rm R}+6{\bar n}_{\rm R}^2 \right).
\label{q}
\end{align}
Its value at $\eta=0$ is a polynomial in the variable ${\bar n}_{\rm R}$:
\begin{align}
& q( {\bar n}_{\rm R}; {\bar n}_{\rm R} )=(1-{\bar n}_{\rm R}^2)(1+2{\bar n}_{\rm R})^2
\left[ 2+6 {\bar n}_{\rm R} +3{\bar n}_{\rm R}^2 (1-{\bar n}_{\rm R} ) \right] .
\label{qas}
\end{align}
The time derivative (\ref{dotG_1}) has the asymptotic behavior:
\begin{align}
& \dot{\cal G}_1(t)=\gamma \, \frac{ q( {\bar n}_{\rm R}; {\bar n}_{\rm R} )
\left[ 2p( {\bar n}_{\rm R}; {\bar n}_{\rm R} )  \right]^{-\frac{1}{2} } }
{ \left[ (1+{\bar n}_{\rm R} ) (1+2{\bar n}_{\rm R} ) \right]^{\frac{5}{2} } }
\, \exp(-\gamma t)            \notag   \\
& +{\rm O}\left[ \exp(-2\gamma t) \right],      \qquad       \qquad     (\gamma t \gg 1).
\label{dotG_1as}
\end{align}
The sign of the vanishing leading term in Eq. (\ref{dotG_1as}) coincides 
with that of the polynomial (\ref{qas}). This polynomial has two positive roots:
${\bar n}_{\rm R}=1 \; \, \text{and} \;\; 
{\bar n}_{\rm R}={\bar n}_{\rm R}^{\prime} \approxeq 2.1022$. The function
$q( {\bar n}_{\rm R}; {\bar n}_{\rm R} )$ is positive outside the interval 
$[1, {\bar n}_{\rm R}^{\prime}]$ and negative inside it. Consequently, the HS evolution 
distance (\ref{G_1}) attains its input-output limit (\ref{G1as}) from below 
if ${\bar n}_{\rm R} <1 \; \, \text{or} \;\; {\bar n}_{\rm R} > {\bar n}_{\rm R}^{\prime}$,   
and from above if ${\bar n}_{\rm R} \in (1, {\bar n}_{\rm R}^{\prime} )$. Therefore, 
we draw the following conclusions.
For ${\bar n}_{\rm R} \leqq 1$ , the function ${\cal G}_1(t)$, Eq. (\ref{G_1}), is monotonic, 
increasing from the initial value ${\cal G}_1(0)=0$ to the asymptotic limit (\ref{G1as}); 
this case is illustrated in Fig. 4. When $1< {\bar n}_{\rm R} <{\bar n}_{\rm R}^{\prime}$, 
then the HS distance  ${\cal G}_1(t)$ evolves non-monotonically, having  a maximum
higher than the equilibrium value (\ref{G1as}), because this is reached from above.
Finally, for ${\bar n}_{\rm R} >{\bar n}_{\rm R}^{\prime}$, either the maximum is followed
by a minimum that is lower than the asymptotic limit (\ref{G1as}), or the function 
${\cal G}_1(t)$ is monotonic. These evolutions of the HS distance  ${\cal G}_1(t)$ 
originate in the analogous ones of the purity ${\cal P}_1(t)$, which are discussed above.

\acknowledgments

This work was supported by the funding agency CNCS-UEFISCDI of the Romanian Ministry 
of Research and Innovation through grant No. PN-III-P4-ID-PCE-2016-0794.


\begin{thebibliography}{100}
\bibitem{HTF2a} Ref. \cite{HTF2}, p. 189: we employ Eqs. (14) and (20) taken 
with the index $\alpha=0$.
\end{thebibliography}
\end{document}